\documentclass[pre,twocolumn,showkeys,showpacs,superscriptaddress,preprintnumbers,floatfix]{revtex4-1}

\usepackage{etex}
\usepackage{ifpdf}
\usepackage{hyperref}
\usepackage{dcolumn}
\usepackage{url}
\usepackage{amsmath}
\usepackage{amscd}
\usepackage{amsfonts}
\usepackage{amssymb}
\usepackage{amsthm}
\usepackage{xfrac}
\usepackage{braket}
\usepackage{bm}   
\usepackage{bbm}
\usepackage{verbatim}
\usepackage{stmaryrd}
\usepackage{xcolor}
\usepackage{setspace}
\usepackage{tikz}
\usetikzlibrary{arrows}
\usetikzlibrary{automata}
\usetikzlibrary{positioning}
\usetikzlibrary{plotmarks}
\usepackage{pgfplots}
\pgfplotsset{compat=newest}

\tikzstyle{vaucanson}=[
  node distance=2.5cm,
  bend angle=15,
  auto,
  every loop/.style={},
  every edge/.style={->,draw=black,line width=1.2,>=latex,shorten <=1pt,shorten >=1pt},
  every state/.style={draw=black,line width=2,font=\large},
  loop right/.style={right,out=22,in=-22,loop},
  loop above/.style={above,out=112,in=68,loop},
  loop left/.style={left,out=202,in=158,loop},
  loop below/.style={below,out=292,in=248,loop},
  loop above right/.style={right,out=67,in=23,loop},
  loop above left/.style={left,out=157,in=113,loop},
  loop below left/.style={left,out=247,in=203,loop},
  loop below right/.style={right,out=337,in=293,loop},
]

\theoremstyle{plain}    
\theoremstyle{plain}    
\theoremstyle{plain}    
\theoremstyle{plain}    
\theoremstyle{plain}    
\theoremstyle{plain}    
\theoremstyle{plain}    
\theoremstyle{plain}    
\theoremstyle{plain}    
\theoremstyle{plain}    
\theoremstyle{plain}    
\theoremstyle{plain}    
\theoremstyle{plain}

\newcommand{\MeasAlphabet}  {\mathcal{A}}

\newcommand{\MeasSymbol}   { {X} }
\newcommand{\meassymbol}   { {x} }

\newcommand{\Past} { \smash{\overleftarrow {\MeasSymbol}} }
\newcommand{\past} { {\smash{\overleftarrow {\meassymbol}}} }

\newcommand{\Future}   { \smash{\overrightarrow{\MeasSymbol}} }
\newcommand{\future}   { \smash{\overrightarrow{\meassymbol}} }

\newcommand{\rep}    { y }
\newcommand{\Rep}    { Y }

\newcommand{\CausalState}   { \mathcal{S} }

\newcommand{\causalstate}   { \sigma }
\newcommand{\CausalStateSet}    { \bm{\CausalState} }
\newcommand{\AlternateState}    { \mathcal{R} }

\newcommand{\AlternateStateSet} { \bm{\AlternateState} }

\newcommand{\Prob}      {\Pr} 

\newcommand{\ProcessAlphabet}   {\MeasAlphabet}

\newcommand{\forward}{+}
\newcommand{\reverse}{-}
\newcommand{\forwardreverse}{\pm} 

\newcommand{\FutureCausalState} { {\CausalState}^{\forward} }

\newcommand{\PastCausalState}   { {\CausalState}^{\reverse} }

\newcommand{\lastindex}[2]{
  \edef\tempa{0}
  \edef\tempb{#2}
  \ifx\tempa\tempb
    \edef\tempc{#1}
  \else
    \edef\tempa{0}
    \edef\tempb{#1}
    \ifx\tempa\tempb
      \edef\tempc{#2}
    \else
      \edef\tempc{#1+#2}
    \fi
  \fi
  \tempc
}

\newcommand{\I}{\mathbf{I}}

\newcommand{\CSjoint}[1][,]{
   \edef\tempa{:}
   \edef\tempb{#1}
   \ifx\tempa\tempb
      \ensuremath{\FutureCausalState\!#1\PastCausalState}
   \else
      \ensuremath{\FutureCausalState#1\PastCausalState}
   \fi
}

\newif\ifpm
\edef\tempa{\forwardreverse}
\edef\tempb{\pm}
\ifx\tempa\tempb
   \pmfalse
\else
   \pmtrue
\fi

\renewcommand{\H}{\operatorname{H}}
\renewcommand{\I}{\operatorname{I}}

\colorlet {R_color}    {blue}
\colorlet {k_color}    {black!30!green}

\def\clap#1{\hbox to 0pt{\hss#1\hss}}

\begin{document}

\title{Prediction and Power in Molecular Sensors:\\
\vspace{0.05in}
Uncertainty and Dissipation When Conditionally Markovian Channels\\ 
Are Driven by Semi-Markov Environments}

\author{Sarah E. Marzen}
\email{semarzen@mit.edu}
\affiliation{Physics of Living Systems, Department of Physics, Massachusetts Institute of Technology,
Cambridge, MA 02139}

\author{James P. Crutchfield}
\email{chaos@ucdavis.edu}
\affiliation{Complexity Sciences Center and Department of Physics, University of
  California at Davis, One Shields Avenue, Davis, CA 95616}

\date{\today}
\bibliographystyle{unsrt}

\begin{abstract}

Sensors often serve at least two purposes: predicting their input and
minimizing dissipated heat. However, determining whether or not a particular
sensor is evolved or designed to be accurate and efficient is difficult. This
arises partly from the functional constraints being at cross purposes and
partly since quantifying the predictive performance of even \textit{in silico}
sensors can require prohibitively long simulations. To circumvent these
difficulties, we develop expressions for the predictive accuracy and
thermodynamic costs of the broad class of conditionally Markovian sensors
subject to unifilar hidden semi-Markov (memoryful) environmental inputs.
Predictive metrics include the instantaneous memory and the mutual information
between present sensor state and input future, while dissipative metrics
include power consumption and the nonpredictive information rate. Success in
deriving these formulae relies heavily on identifying the environment's causal
states, the input's minimal sufficient statistics for prediction. Using these
formulae, we study the simplest nontrivial biological sensor model---that of a
Hill molecule, characterized by the number of ligands that bind simultaneously,
the sensor's cooperativity.  When energetic rewards are proportional to total
predictable information, the closest cooperativity that optimizes the total
energy budget generally depends on the environment's past hysteretically. In
this way, the sensor gains robustness to environmental fluctuations. Given the
simplicity of the Hill molecule, such hysteresis will likely be found in more
complex predictive sensors as well. That is, adaptations that only locally
optimize biochemical parameters for prediction and dissipation can lead to
sensors that ``remember'' the past environment.


\end{abstract}

\noindent

\preprint{Santa Fe Institute Working Paper 2017-07-XXX}
\preprint{arxiv.org:1707.XXXX [physics.gen-ph]}

\maketitle


\setstretch{1.1}

\newcommand{\Abet}{\ProcessAlphabet}
\newcommand{\MS}{\MeasSymbol}
\newcommand{\ms}{\meassymbol}
\newcommand{\SSet}{\CausalStateSet}
\newcommand{\St}{\CausalState}
\newcommand{\st}{\causalstate}
\newcommand{\MxSt}{\AlternateState}
\newcommand{\MxSSet}{\AlternateStateSet}
\newcommand{\mxst}{\mu}
\newcommand{\mxstt}[1]{\mu_{#1}}
\newcommand{\StartMS}{\bra{\delta_p}}
\newcommand{\FSt}{\St^+}
\newcommand{\fst}{\st^+}
\newcommand{\PSt}{\St^-}
\newcommand{\pst}{\st^-}
\newcommand{\ChanAlph}{\mathcal{Y}}
\newcommand{\Tau}{\mathcal{T}}


\vspace{0.2in}

\paragraph*{Introduction}
To perform functional tasks, synthetic nanoscale machines and their
macromolecular cousins simultaneously manipulate energy, information, and
matter. They are \emph{information engines}---systems that operate by
synergistically balancing the energetics of their physical substrate against
required information generation, storage, loss, and transformation to support a
given functionality. Classically, information engines were conceived as either
potential computers \cite{bennett1982thermodynamics}---that is, physical
systems that can compute anything given the right program---or as
Maxwellian-like demons that use information as a resource to convert disordered
energy to useful work
\cite{Maxw88a,Szil29a,mandal2012work,deffner2013information}. Recently,
investigations into functional computation \footnote{Here, when analyzing
sensory information processing in biological systems, we take care to
distinguish intrinsic, functional, and useful computation
\cite{Crut88a,Crut92c,Crutchfield&Mitchell94a}. \emph{Intrinsic computation}
refers to how a physical system stores and transforms its historical
information. We take \emph{functional computation} as information processing in
a physical device that promotes the performance of a larger, encompassing
system. Whereas, we take \emph{useful computation} as information processing in
a physical device used to achieve an external user's goal. The first is
well-suited to analyzing structure in physical processes and determining if
they are candidate substrates for any kind of information processing. The
second is well-suited for discussing biological sensors, while the third is
well-suited for discussing the benefits of contemporary digital computers.}
embedded in physical systems led to studies of the thermodynamics of various
kinds of information processing \cite{parrondo2015thermodynamics}, including
the thermodynamic costs of information creation \cite{Agha16d}, noise
suppression \cite{hinczewski2014cellular}, error correction and synchronization
\cite{Boyd16c}, prediction \cite{Still2012,barato2014efficiency,Horowitz2014},
homeostasis \cite{Boyd16d}, learning \cite{goldt2017stochastic}, structure
\cite{Boyd16e}, and intelligent control \cite{Boyd14b}.

Due to its broad importance to the survival of biological organisms, here we
focus on a specific functional computation in information engines: how sensory
subsystems predict their environment. And, we introduce a thermodynamic
analysis that can address environmental processes more complex than the
memoryless and finite-state Markov sources and Gaussian processes considered in the above cited works.

Evolved and designed sensory systems are often tasked with at least two,
potentially competing, objectives: accurately predicting inputs
\cite{rao1999predictive, Palm13a} and minimizing required power and heat
dissipation \cite{chklovskii2004maps, hasenstaub2010metabolic} \footnote{We
only consider online or real-time computations, so that the oft-considered
energy-speed-accuracy tradeoff \cite{lan2012energy, lahiri2016universal}
reduces to an energy-accuracy tradeoff.}. Accurate and energy-efficient
predictive models can be used to optimize action policies, for example, to reap
increased rewards from the environment, even regardless of one's particular
``reward function'' \cite{sutton1998reinforcement, littman2001predictive,
Brodu11, little2014learning}.

Extracting such models from actual biological sensors requires a nontrivial
matching of sensory statistics and sensor structure \cite{becker2015optimal,
Boyd16d}. Undaunted, much effort has been invested to find minimally
dissipative and maximally predictive sensors, often simplifying the challenges
by ignoring action policies---how the sensed information is used. Some seek
sensor models that maximize a combination of predictive power and (energetic)
efficiency; e.g. as in Refs. \cite{barato2014efficiency, becker2015optimal}.
Others validate learning rules based on whether or not they maximize the
aforementioned objective function \cite{creutzig2008predictive,
creutzig2009past}. Finally, others compare real biological sensors to \emph{in
silico} null models; e.g. as in Ref. \cite{Palm13a}.

These efforts require calculating predictive and dissipative metrics of sensory
models. 
For realistic
null models, estimating predictive power from simulation can
require prohibitively long simulations. To circumvent these and related
difficulties, then, one desires closed-form expressions for various predictive
and dissipative metrics in terms of the generators of environment behavior and
of the sensor-channel properties---the input-dependent stochastic dynamical
system describing the sensor. The following provides these expressions for
quite general sensors: conditionally Markovian channels subject to complex
nonequilibrium steady-state environments that are unifilar hidden semi-Markov
processes. Our derivations rely heavily on a recent characterization of the
minimal sufficient statistics for predicting such complex environments
\cite{Marz17b}.

\begin{figure}
\centering
\includegraphics[width=\columnwidth]{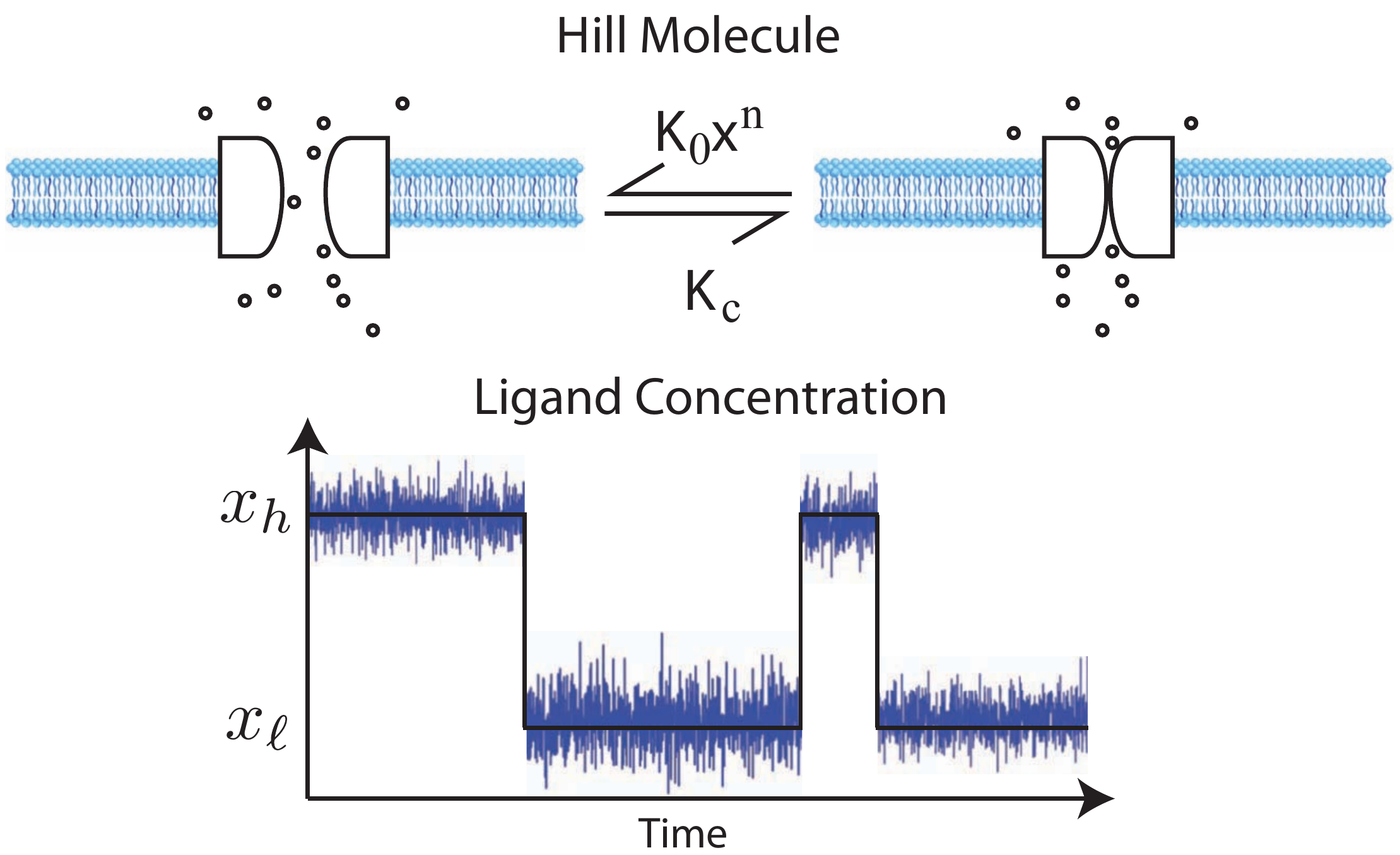}
\caption{Hill molecule: (Top) Ion channel in a neuronal membrane in the open
	(left) and closed (right) states in which ions can and cannot travel
	through. (Below) Ligand concentrations during the open-closed cycle of the
	molecule.
    }
\label{fig:HillMoleculeConception}
\end{figure}

To illustrate the insights that arise from this, we study an ``optimal'' Hill
molecule model of ligand-gated channels; see Fig.
\ref{fig:HillMoleculeConception}. A Hill molecule is a conditionally Markovian
channel with its ligand concentration as input. We assume that the ligand
concentration is a realization of a semi-Markov process, a generalization over
previous efforts that assumed Markovian \cite{barato2014efficiency} or Gaussian
\cite{becker2015optimal} processes. Our generalization to arbitrarily
temporally correlated inputs is necessary when, for instance, the Hill molecule
represents a nicotinic acetylcholine receptor on a synapse and ligands are
acetylcholine molecules since, as a practical matter, neuronal dynamics are
often non-Markovian and non-Gaussian \cite{izhikevich2007dynamical}. As a
result, we find that (i) increases in cooperativity (memory) of the Hill
molecule lead to increases in both predictive power and power consumption, (ii)
a large fraction of power consumption comes from inefficient prediction, and
(iii) simple gradient-based adaptation rules lead to hysteresis.

\paragraph*{\textbf{Background}}
\label{sec:Background}
Central to our analysis is an appreciation of causal states (minimal
sufficient statistics of prediction and/or retrodiction), unifilar hidden
semi-Markov processes, and conditionally Markovian channels. We review these
concepts here, simultaneously introducing relevant notation.

\paragraph*{Environment}
Input symbols $\ms$ take on any value in the observation alphabet
$\MeasAlphabet$. We code the \emph{past} of the input time series as
$\overleftarrow{\ms} = \ldots
(\ms_{-2},\tau_{-2}),(\ms_{-1},\tau_{-1}),(\ms_0,\tau_+)$ and the input's
\emph{future} as $\overrightarrow{\ms} =
(\ms_0,\tau_-),(\ms_1,\tau_1),(\ms_2,\tau_2),\ldots$, where $\tau_i$ is the
total dwell time for symbol $\ms_i$. To ensure a unique coding, we stipulate
that $\ms_i\neq\ms_{i+1}$. Note that symbol $\ms_0$ is seen for a total dwell
time of $\tau_+ + \tau_-= \tau_0$; that is, the present splits the dwell time
$\tau_0$ into two.

As is typical, $\Past$ is the random variable corresponding to semi-infinite
input pasts and $\Future$ the random variable corresponding to semi-infinite
input futures. We now briefly review the definition of causal states, as
described in Ref. \cite{Shal98a}. Forward-time causal states $\FSt$, the
minimal sufficient statistics for prediction, are defined via the following
equivalence relation: two semi-infinite pasts, $\past$ and $\past'$, are
considered ``predictively'' equivalent if:
\begin{align*}
\past\sim_{\epsilon^+}\past'
  \Leftrightarrow \Prob(\Future|\Past=\past) = \Prob(\Future|\Past=\past')
  ~.
\end{align*}
The relation partitions the set of semi-infinite pasts into clusters of pasts.
Each cluster is a \emph{forward-time causal state} $\fst$. \emph{Reverse-time
causal states} $\PSt$, the minimal sufficient statistics for retrodiction, are
defined similarly. Two semi-infinite futures, $\future$ and $\future'$, are
considered ``retrodictively'' equivalent if:
\begin{align*}
\future\sim_{\epsilon^-}\future'
  \Leftrightarrow \Prob(\Past|\Future=\future) = \Prob(\Past|\Future=\future')
  ~.
\end{align*}
This equivalence relation partitions the set of semi-infinite futures into
clusters, each cluster being a reverse-time causal state $\pst$.

Forward- and reverse-time causal states are useful in the ensuing calculations
due to the following Markov chains. First, forward-time causal states are a
deterministic function of the input past ($\fst = \epsilon^+(\past)$), and
reverse-time causal states are a deterministic function of the input future
($\pst = \epsilon^-(\future)$). Hence, we have the Markov chains
$\FSt\rightarrow\Past\rightarrow\Future$ and
$\PSt\rightarrow\Future\rightarrow\Past$. However, causal states are minimal
sufficient statistics of the past relative to the future and vice versa. And
so, $\Past\rightarrow\FSt\rightarrow\Future$ and
$\Future\rightarrow\PSt\rightarrow\Past$ are also valid Markov chains.
Invoking these Markov chains is called \emph{causal shielding}.

Let's first address the more general case of unifilar hidden semi-Markov input,
as in Ref. \cite{Marz17b}. Forward-time hidden states are labeled
$g$, and causal states are thus labeled by $(g,\ms_+,\tau_+)$. That is, the
forward-time hidden state $g$, current emitted symbol $\ms_+$, and time since
last symbol $\tau_+$ together comprise the forward-time causal states for
unifilar hidden semi-Markov input processes. Dwell times are drawn from
$\phi_g(\tau)$; emitted symbols are chosen with probability $p(\ms|g)$; and
$g = \epsilon^+(g',\ms')$ is the next hidden state given that the current
hidden state is $g'$ and the current emitted symbol is $\ms'$.

For the Hill molecule, we focus on semi-Markov input. This greatly constrains
the forward- and reverse-time causal states, so that $g$ and $\ms_+$ are
equivalent. The forward-time causal states are thus described by the pair
$(\ms_+,\tau_+)$, where $\ms_+$ is the input symbol infinitesimally prior to
the present and $\tau_+$ is the time since last symbol (i.e., $\ms_{-1}$). The
reverse-time causal states are similarly described by the pair
$(\ms_-,\tau_-)$, where $\ms_-$ is the input symbol infinitesimally after the
present and $\tau_-$ is the time to next symbol (i.e., $\ms_1$). Let
$\Tau_{\pm}$ be the random variable describing time since (to) last (next)
symbol. The dwell time of symbol $\ms$ has probability density function
$\phi_{\ms}(\tau)$, and the probability of observing symbol $\ms$ after $\ms'$
is $q(\ms|\ms')$. By virtue of how we have chosen to encode our input: $q(\ms|\ms)=0$.

Finally, the development to come requires our finding the joint distribution
$\rho(\fst,\pst)$ of forward- and reverse-time causal states. When the input is
unifilar hidden semi-Markov, efficiently finding $\rho(\fst,\pst)$ is an open
problem. Nonetheless, we can say:
\begin{align*}
\rho(\pst|\fst)
  & = \rho(g_-,\ms_-,\tau_-|g_+,\ms_+,\tau_+) \\
  & =\delta_{\ms_+,\ms_-}
  \frac{\phi_{g_+}(\tau_+ +\tau_-)}{\Phi_{g_+}(\tau_+)} p(g_+|g_-,\ms_-)
  ~.
\end{align*}
For semi-Markov input, this
simplifies: $\rho(\fst,\pst) = \rho((\ms_+,\tau_+),(\ms_-,\tau_-))$.  As
described in Ref. \cite{Marz17b}, we have:
\begin{align}
\rho(\ms_+,\tau_+) = \mu_{\ms_+} \Phi_{\ms_+}(\tau_+) p(\ms_+)
\label{eq:marg_pdf}
\end{align}
where:
\begin{align*}
\Phi_{\ms_+}(\tau_+) = \int_{\tau_+}^{\infty} \phi_{\ms_+}(t) dt ,~
\mu_{\ms_+} = 1/\int_0^{\infty} t\phi_{\ms_+}(t)dt~,
\end{align*}
and $p(\ms_+)$ is the probability of observing symbol $\ms_+$.

The latter probability is given by:
\begin{align*}
p(\ms_+) = \left( \text{diag}(1/\mu_{\ms}) ~\text{eig}_1(q) \right)_{\ms_+}
  ~.
\end{align*}
The conditional distribution of reverse-time causal states given forward-time
causal states is then:
\begin{align}
\rho(\pst|\fst)
  & = \rho((\ms_-,\tau_-)|(\ms_+,\tau_+)) \nonumber \\
  & = \frac{\phi_{\ms_+}(\tau_+ + \tau_-)}{\Phi_{\ms_+}(\tau_+)}
  \delta_{\ms_+,\ms_-}
  ~.
\label{eq:cond_pdf}
\end{align}
Together, Eqs.~(\ref{eq:marg_pdf}) and (\ref{eq:cond_pdf}) give the joint distribution $\rho(\fst,\pst)=\rho(\fst)\rho(\pst|\fst)$.

\paragraph*{Sensory channel}
We assume the channel is conditionally Markovian. As such, its dynamics are
fully specified by input state-dependent kinetic rates. More precisely, the
channel state $\rep$, with corresponding random variable $\Rep$, can take on
any value in $\ChanAlph$, and the rate at which channel state $\rep$ transfers
to channel state $\rep'$ when the input has value $\ms$ is given by:
\begin{align*}
k_{\rep\rightarrow\rep}(\ms) := -\sum_{\rep'} k_{\rep\rightarrow\rep'}(\ms)
  ~.
\end{align*}
Then, the probability $p(\rep,t)$ of being in channel state $\rep$ at time $t$
evolves as:
\begin{align*}
\frac{dp(\rep,t)}{dt} = \sum_{\rep'} k_{\rep'\rightarrow\rep}(\ms(t)) p(\rep',t)
  ~,
\end{align*}
where $\ms(t)$ is the input symbol at time $t$.

To simplify notation and ease computation, we write dynamical evolution rules
in matrix-vector form. Let $\vec{p}(\rep,t)$ be the vector of probabilities
that the channel is in a particular state $\rep$ at time $t$, and let $M(\ms)$
be a matrix of rates: $M_{\rep',\rep}(\ms) = k_{\rep\rightarrow\rep'}(\ms)$.
Then, we have:
\begin{align}
\frac{d\vec{p}(\rep,t)}{dt} = M(\ms(t)) \vec{p}(\rep,t)
  ~.
\label{eq:channel_evo}
\end{align}
With this, it is clear that $\vec{p}(\rep,t)$ can oscillate or decay to a
steady state. The Perron-Frobenius theorem guarantees that:
\begin{align*}
p_{eq}(\ms) := \text{eig}_0(M(\ms))
  ~,
\end{align*}
the probability distribution over channel states when ligand concentration is
set to $\ms$, is unique.

\paragraph*{\textbf{Sensor Accuracy and Thermodynamics}}
We now introduce and justify predictive and dissipative metrics, present
closed-form expressions for these metrics in terms of aforementioned generators
and channels, and explore the relationship between cooperativity in biochemical
sensing and prediction and dissipation.

\newcommand{\ProtFuture}{\protect\overrightarrow{X}}
\newcommand{\Imem}{\I_\text{mem}}
\newcommand{\Ifut}{\I_\text{fut}}
\newcommand{\Inpdot}{\dot{\I}_\text{np}}

\begin{figure}
\centering
\includegraphics[width=\columnwidth]{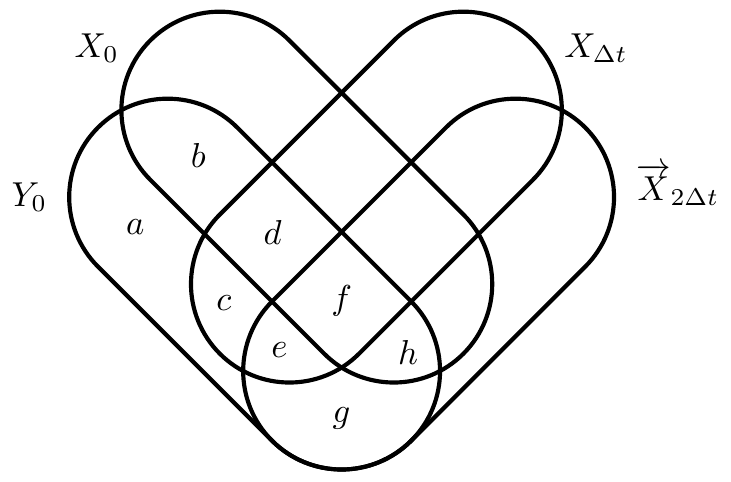}
\caption{Sensor information diagram \cite{Jame11a} giving the relationship
	between predictive metrics and nonpredictive information rate.
	Instantaneous memory is $\Imem = \I[X_0;Y_0] = b+d+f+h$ and total
	predictable information $\Ifut = \lim_{\Delta t\rightarrow 0}
	\I[Y_0;X_{0:}] = \lim_{\Delta t\rightarrow 0} b+c+d+e+f+g+h$, while the
	lower bound on power consumption, the nonpredictive information rate is
	$\Inpdot = \lim_{\Delta t\rightarrow 0} \frac{b+h-c-e}{\Delta t}$.
	Recall $a = \H[Y_0|X_0,X_{\Delta t},\ProtFuture_{2 \Delta t}]$,
	$b = \I[Y_0;X_0|X_{\Delta t},\ProtFuture_{2 \Delta t}]$,
	$f = \I[Y_0;X_0;X_{\Delta t};\ProtFuture_{2 \Delta t}]$, and so on.
    }
\label{fig:iDiagram}
\end{figure}

\paragraph*{Predictive and dissipative metrics}
We employ several metrics to quantify the sensor predictive performance and
their energy efficiency. \emph{Instantaneous memory} $\Imem$ \cite{Still2012}
and \emph{total predictable information} $\Ifut$ \cite{Stil07b,
creutzig2009past} characterize predictive power, while the \emph{nonpredictive
information rate} $\Inpdot$ \cite{Still2012, Horowitz2014,
barato2014efficiency} and temperature-normalized \emph{power consumption}
$\beta P$ monitor dissipation. Figure \ref{fig:iDiagram} uses an information
diagram \cite{Yeun91a,Jame11a} to illustrate their relations in terms of
the elementary information atoms---entropies, conditional entropies, and
mutual informations---out of which they are constructed.

Our selection of metrics differs from previous efforts to characterize
prediction and dissipation. For instance, the instantaneous predictive
information in Ref. \cite{Still2012} is equivalent to instantaneous memory to
$O(\Delta t)$ in a continuous-time framework. Similarly, Ref.
\cite{Horowitz2014} focused on instantaneous memory and nonpredictive
information rate, but did not calculate total predictable information or a more
standard prediction-related metric. Reference \cite{barato2014efficiency} used
the ratio of nonpredictive information rate to entropy production to
characterize learning, but nonpredictive information rate is not a typical
metric for predictive power in machine learning or related literature. Finally,
Ref. \cite{becker2015optimal} focused on metrics for prediction, including a
natural continuous-time extension of instantaneous predictive information, but
not on metrics for dissipation.


There is little consensus on quantifying a channel's predictive capability.
Common metrics are designed to quantify \emph{memory} rather than
\emph{prediction} \cite{white2004short}, but even when adapted for measuring
prediction, one can choose different types of readout function. We focus on
what we call the \emph{total predictable information}:
\begin{align*}
\Ifut := I[\Rep;\Future]
  ~.
\end{align*}
This is the mutual information between present channel state and the input's
future. It is the amount of information that is predictable about the input
future from the present channel state. Due to the feedforward nature of the
channel-input setup---that is, the channel's state does not affect the
input---and the Markov chains given earlier, we have the Markov chain
$\Rep\rightarrow\FSt\rightarrow\PSt\rightarrow\Future$. As a result: \begin{align*}
\Ifut = \I[\Rep;\PSt] = \I[\Rep;\MS_-,\Tau_-]
  ~,
\end{align*}
which decomposes into:
\begin{align*}
\Ifut = \I[\Rep;\MS_-] + \I[\Rep;\Tau_-|\MS_-]
  ~.
\end{align*}
The term $\I[\Rep;\MS_-]$ is called the \emph{instantaneous memory} $\Imem$
\cite{Still2012}, since it is the amount of information available from the
channel state about the just-seen input symbol. We find that:
\begin{align*}
\Ifut = \Imem + \I[\Rep;\Tau_-|\MS]
  ~.
\end{align*}
Thus, the total predictable information is the sum of instantaneous memory and
information that is truly about the future, which here is the time to next
symbol.

The Supplementary Material justifies $\Ifut$ on generational timescales as a metric via an extension of
Kelly's classic bet-hedging argument \cite{Cove06a}. In a discrete-time
setting, increases in growth rate via increases in sensory information is equal
to the instantaneous predictable information $\I[\Rep_0;\MS_{\Delta t}]$. The
total predictable information $\Ifut$ is an upper bound on this increase in
growth rate.  On ontogenetic timescales, we merely assert that total predictable information might increase concurrently with energetic rewards.


Next, we need to quantify the power consumed by the sensor system. Assuming
access to a temperature-normalized ``energy function'' $\beta E(\ms,\rep)$,
the temperature-normalized power $\beta P$ is given by:
\begin{align}
\beta P
  & = \lim_{\Delta t\rightarrow 0}
  \frac{\langle \beta E(\ms_{t+\Delta t},\rep_t)\rangle -\langle \beta E(\ms_t,\rep_t) \rangle}{\Delta t}
  ~.
\label{eq:betaP0}
\end{align}
If determining an energy function is not possible, we can calculate a lower
bound using a continuous-time adaptation of the inequality in Ref.
\cite{Still2012}:
\begin{align}
\Inpdot := \lim_{\Delta t\rightarrow 0}
  \frac{\I[\Rep_t;\MS_t]-\I[\Rep_t;\MS_{t+\Delta t}]}{\Delta t} \leq \beta P
  ~,
\label{eq:ToP}
\end{align}
with an alternate equivalent definition in Ref. \cite{Horowitz2014}; see the
Supplementary Material.

$\Inpdot$ is called the \emph{nonpredictive information rate} since it loosely
corresponds to how much of the instantaneous memory is useless for predicting
the next input. Reference \cite{barato2014efficiency} viewed $\Inpdot/\beta P$
as a learning efficiency. We take the view that $\Inpdot$ is a potentially
useful lower bound on temperature-normalized power consumption, and use $\Imem$ and $\Ifut$
instead to characterize learning. Any differences between the formulae shown
here and Ref. \cite[Eq. (2)]{Still2012} are superficial; we merely adapted the
derivation for continuous-time processes. Unfortunately, no one has yet given a
guarantee that the nonpredictive information rate is a tight lower bound on
temperature-normalized power.

\paragraph*{Closed-form metrics}

As stated earlier, we wish to find closed-form expressions for $\Ifut$,
$\Imem$, $\Inpdot$, and $\beta P$ in terms of input
properties---$\phi_{\ms}(\tau)$, $\epsilon^+(g,\ms)$, $p(\ms|g)$---and channel
properties---$M(\ms)$. Their derivations are too lengthy for here and so are
relegated to the Supplementary Materials (SM). The appropriate equations there
are referenced here, where relevant.


Calculating $\Ifut$ and $\Imem$ can be accomplished once $\rho(\fst,\rep) =
\Prob(\FSt=\fst,\Rep=\rep)$ is obtained.  This follows, in turn, by
manipulating a Chapman-Kolmogorov equation, shown in the SM.  Set any ordering
on the pairs $(g,\ms)$; e.g., the ordering
$(g_1,\ms_1),(g_1,\ms_2),\ldots,(g_{|\mathcal{G}|},\ms_{|\MeasAlphabet|})$. An
expression for $p(\rep|\fst) = p(\rep|g,\ms,\tau)$ is given by a combination of
Eqs.~(\ref{eq:5}) and (\ref{eq:init_conds}):
\begin{align*}
p(\rep|g,\ms,\tau)
  = \left(
  e^{M(\ms)\tau} \text{eig}_1(\textbf{C})_{(g,\ms)}/\mu_{g} p(g)
  \right)_{\rep}
  ~,
\end{align*}
where $\textbf{C}$ is a block matrix with entries
$\textbf{C}_{(g,\ms),(g',\ms')} = \delta_{g,\epsilon^+(g',\ms')} p(\ms'|g')
\int_0^{\infty} \phi_{g'}(t) e^{M(\ms')t} dt$. And so,
$\text{eig}_1(\textbf{C})$ is a block vector. Normalization forces
$\vec{1}^{\top} \text{eig}_1(\textbf{C})_{(g,\ms)} = \mu_{g} p(g)$.

We then find the joint probability distribution as $\rho(\rep,\fst) =
p(\rep|\fst) \rho(\fst)$, which enables computation of all predictive metrics.
Instantaneous memory is given by  $\Imem = \I[\MS;\Rep]$, whereas $\Ifut =
\I[\Rep;\PSt]$. All the relevant distributions---namely, $p(\ms,\rep)$ and
$\rho(\pst,\rep)$---are obtained from the previously derived
$\rho(\fst,\rep)$.  For instance, to calculate $\rho(\pst,\rep)$, we employ the
previously stated Markov chain to find:
\begin{align*}
\rho(\pst,\rep) = \sum_{\fst} \rho(\pst|\fst) p(\rep|\fst) \rho(\fst)
  ~.
\end{align*}
And, to calculate $p(\ms,\rep)$, we recall that $\pst = (\ms,\tau_-)$, so we
only need marginalize the joint distribution of $\rho((\ms,\tau_-),\rep)$.


Calculation of dissipative metrics can be additionally accomplished once:
\begin{align*}
\frac{\delta p(\ms,\rep)}{\delta t}
  & = \lim_{\Delta t\rightarrow 0}
  \Big(\Prob(\MS_{t+\Delta t}=\ms,\Rep_t=\rep) \nonumber \\
  & \quad - \Prob(\MS_t=\ms,\Rep_t=\rep) \Big)/\Delta t
\end{align*}
is obtained. An expression for $\frac{\delta p}{\delta t}$ in terms of input
and channel properties is given in Eq.~(\ref{eq:deltap_deltat1}):
\begin{align*}
\frac{\delta p}{\delta t}
  & = \sum_{g',\ms'\neq\ms}
  \int d\tau'
  ~p(\ms|\epsilon^+(g',\ms')) p(\ms'|g') \phi_{g'}(\tau') \nonumber \\
  & \quad \times \left(e^{M(\ms')\tau'}
  \text{eig}_1(\textbf{C})_{(g',\ms')}\right)_{\rep} \nonumber \\
  & \quad - \sum_{g'} \int d\tau' ~p(\ms|g') \phi_{g'}(\tau')
  \Big(e^{M(\ms)\tau'} \text{eig}_1(\textbf{C})_{(g',\ms)}\Big)_{\rep}
  ,
\end{align*}
where normalization again requires $\vec{1}^{\top} \text{eig}_1(\textbf{C})_{(g,\ms)} = \mu_{\ms} p(g)$. Then, from earlier, we find that:
\begin{align}
\Inpdot & = \lim_{\Delta t\rightarrow 0}
  \frac{H[\Rep_t,\MS_{t+\Delta t}]-H[\Rep_t,\MS_t]}{\Delta t} \\
  & = \lim_{\Delta t\rightarrow 0}
  \Big(
  \sum_{\ms,\rep} \left(p(\ms,\rep)+\frac{\delta p}{\delta t}\Delta t\right)
  \log \frac{1}{p(\ms,\rep)+\frac{\delta p}{\delta t}\Delta t}\nonumber \\
  & \qquad - \sum_{\ms,\rep} p(\ms,\rep) \log \frac{1}{p(\ms,\rep)}\Big)/\Delta t \\
  & = -\sum_{\ms,\rep} \frac{\delta p(\ms,\rep)}{\delta t} \log p(\ms,\rep)
  ~.
\label{eq:Inps}
\end{align}
When there are no nondecaying oscillations in $\vec{p}(\rep,t)$ \cite{schnakenberg1976network}, we can find $\beta P$ despite lacking direct access to an energy function by calculating the steady-state distribution over channel states with fixed input:
\begin{align*}
\vec{p}_{eq}(\rep|\ms)
  = \text{eig}_0(M(\ms)) = \frac{e^{-\beta E(\ms,\rep)}}{Z_{\beta}(\ms)}
  ~,
\end{align*}
where the partition function is $Z_{\beta}(\ms) := \sum_{\rep} e^{-\beta E(\ms,\rep)}$. Hence:
\begin{align}
\beta E(\ms,\rep)
  = \log \frac{1}{p_{eq}(\rep|\ms)} - \log Z_{\beta}(\ms)
  ~.
\label{eq:energy_function}
\end{align}
Recalling Eq.~(\ref{eq:betaP0}) and invoking stationarity---that
$\Prob(\MS_t=\ms)=\Prob(\MS_{t+\Delta t}=\ms)$---yields:
\begin{align}
\beta P
  & = \lim_{\Delta t\rightarrow 0}
  \Big( \langle \log\frac{1}{p_{eq}(\ms,\rep)}
  \rangle_{\Prob(\MS_{t+\Delta t}=\ms,\Rep_t=\rep)} \nonumber \\
  & \qquad - \langle \log\frac{1}{p_{eq}(\ms,\rep)}
  \rangle_{\Prob(\MS_t=\ms,\Rep_t=\rep)}\Big)/\Delta t \\
  & = \sum_{\ms,\rep}
  \frac{\delta p(\ms,\rep)}{\delta t} \log \frac{1}{p_{eq}(\rep|\ms)}
  ~.
\label{eq:betaP}
\end{align}
The distributions $\Prob(\MS_{t+\Delta t}=\ms,\Rep_t=\rep)$ and
$\Prob(\MS_t=\ms,\Rep_t=\rep)$ can be obtained from $M(\ms)$. In other words,
when there are no recurrent cycles, we can calculate $\beta P$ directly from
the kinetic rates $k_{\rep\rightarrow\rep'}(\ms)$ and input generator
($\phi_g(\tau),~\epsilon^+(g,\ms),~p(\ms|g)$) alone.


\paragraph*{Effect of cooperativity on prediction and dissipation}
The Hill molecule is a common fixture in theoretical biology, as it is the
simplest mechanistic model of cooperativity \cite{marzen2013statistical}.
Recall Fig.~\ref{fig:HillMoleculeConception}. A Hill molecule can be in one of
two states, open or closed. When open, $n$ ligand molecules are bound; when
closed, no ligand molecules are bound. Hence, the state of the Hill molecule
carries information about the number of bound ligand molecules. In other words,
such molecules are sensors of their environment.

Let us outline a simple dynamical model of the Hill molecule. The rate of
transition from closed $C$ to open $O$ given a ligand concentration $x$ is:
\begin{align}
k_{C\rightarrow O} = k_O x^n
  ~.
\label{eq:Hill1}
\end{align}
While the transition rate from open $O$ to closed $C$ is:
\begin{align}
k_{O\rightarrow C} = k_C
  ~.
\label{eq:Hill2}
\end{align}
The steady-state distribution given fixed ligand concentration is the familiar:
\begin{align*}
\text{Pr}_{\text{eq}} (\Rep=O|\MS=\ms) & = \frac{k_O x^n}{k_C + k_O x^n} \\
                           & = \frac{x^n}{(k_C/k_O) + x^n}
  ~.
\end{align*}
Although the mechanistic model makes sense only when $n$ is a nonnegative
integer, this model is often used when $n$ is any nonnegative real number;
increases in $n$ can still be thought of as increases in cooperativity.
Equations~(\ref{eq:Hill1})-(\ref{eq:Hill2}) constitute a complete
characterization of channel properties.

Increasing the cooperativity $n$ increases the steepness of the molecule's ``binding
curve''---the probability of being ``on'' as a function of concentration. In
other words, the sensor becomes more switch-like and less a proportionately
responding transducer of the input. If the concentration is greater than $\left(k_C/k_O\right)^{1/n}$,
the switch is essentially ``on'' if $n$ is high. A more switch-like sensor is
useful if the optimal phenotype depends only upon the condition ``ligand
concentration greater than X''. Whereas, a less switch-like, smoother
responding sensor helps if the optimal phenotype depends on ligand
concentration in a more graded manner.

The concentration scale is set by $\left(k_C/k_O\right)^{1/n}$, while the time
scale is set by $1/k_C$; as such, we set both to $k_O=k_C=1$ without loss of
generality. We imagine that the ligand concentration alternates between two
values: $\ms_l$ and $\ms_h$. When there is less ligand ($\ms_l$), we will tend
to see the Hill molecule revert to and stay in the closed state.  When there is
ligand ($\ms_h>\ms_l$), we will tend to see the Hill molecule revert to and
stay in the open state. With no particular application in mind, we imagine
that the dwell-time distributions take the form $\phi_{\ms}(\tau) =
\lambda(\ms)^2\tau e^{-\lambda(\ms)\tau}$ with $\lambda(\ms_l)=5$ and
$\lambda(\ms_h)=4$.

\begin{figure}
\centering
\includegraphics[width=\columnwidth]{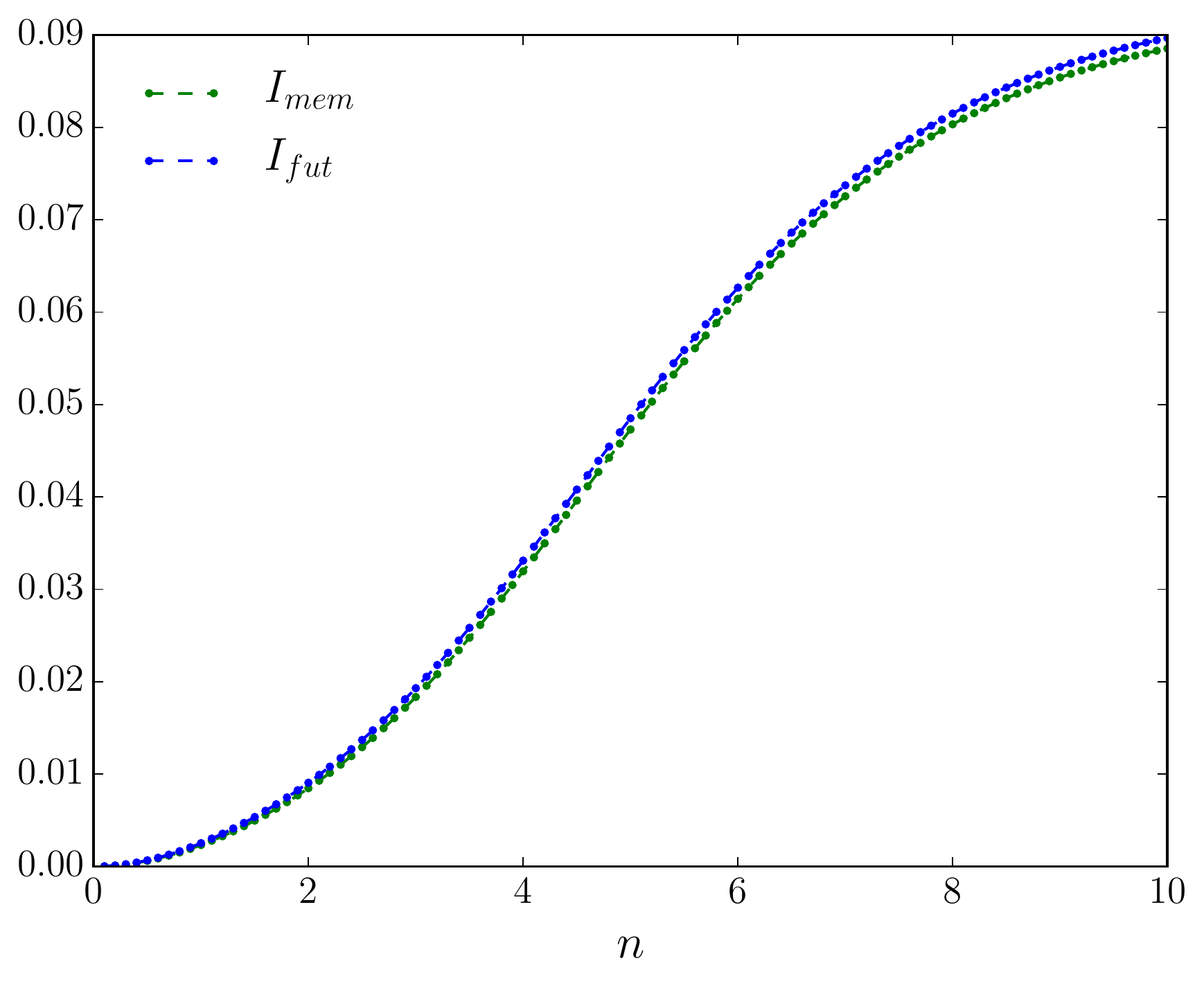}
\includegraphics[width=.96\columnwidth]{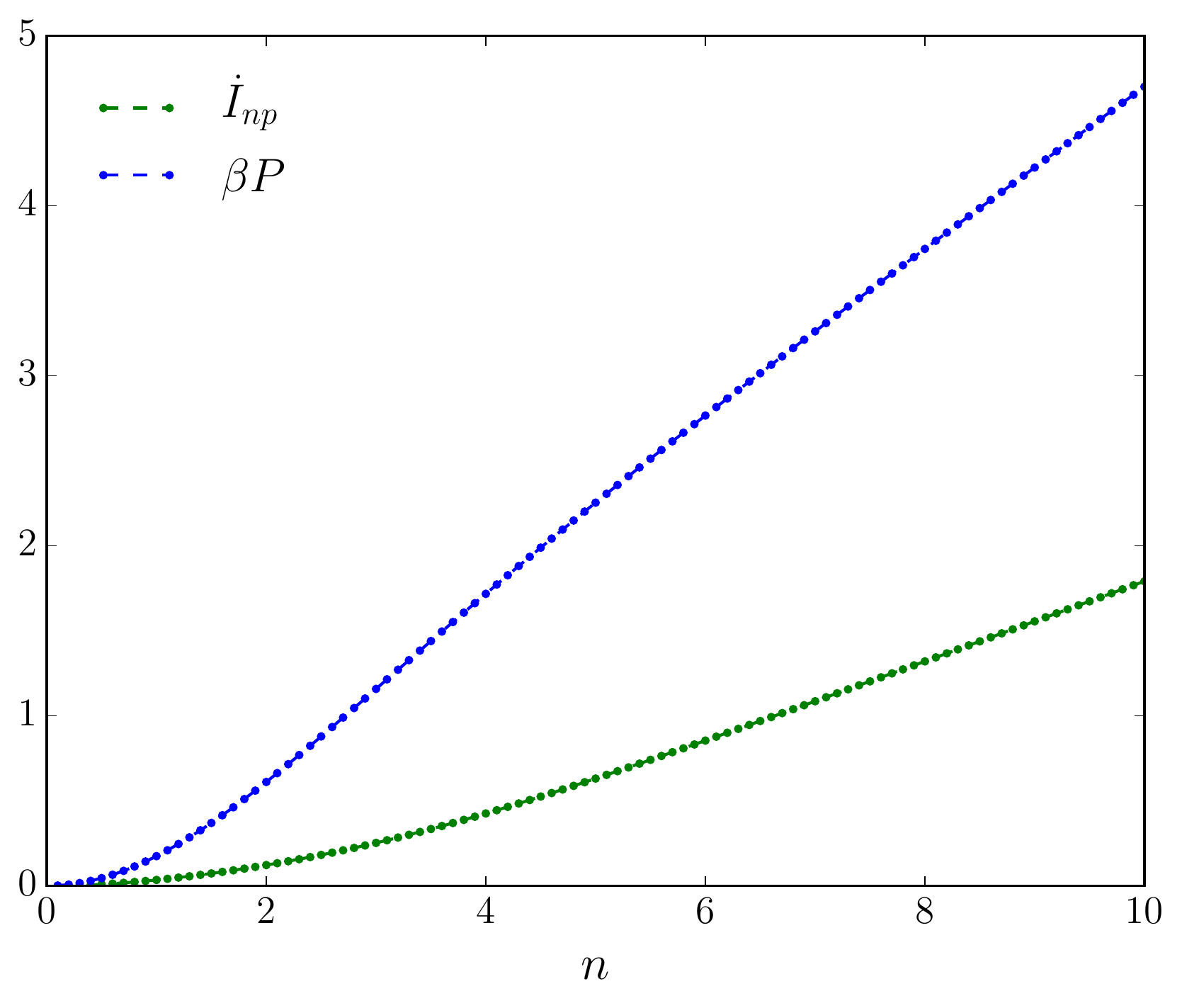}
\caption{(Top) Ligand concentration can take one of two values, $\ms_l = 0.5$
	and $\ms_h = 2.0$, in units of $\left(k_C/k_O\right)^{1/n}$.  Dwell-time
	distributions are parametrized as $\phi_{\ms}(t) = \lambda(\ms)^2 t
	e^{-\lambda(\ms)t}$ with $\lambda(\ms_l) = 5,~\lambda(\ms_h)=4$ in units of
	$1/k_C$.  Hill molecule parameters are set to $k_O=k_C=1$, with varying
	cooperativity $n$. (Bottom) prediction-related metrics: instantaneous
	memory $I_{mem}$ and total predictive power $I_{fut}$, as functions of $n$
	in nats.  At right, temperature-normalized dissipation-related metrics:
	non-predictive information rate $\dot{I}_{np}$ and temperature-normalized
	power $\beta P$, both in nats per unit time.
	}
\label{fig:hill}
\end{figure}

We now deploy the earlier formulae to study the predictive capabilities and
dissipative tendencies of a Hill molecule subject to semi-Markov input.
Previous studies of biological sensors found that increases in cooperativity
accompanied increases in channel capacity \cite{tkavcik2009optimizing,
walczak2010optimizing, martins2011trade}. Others studied the thermodynamics of
prediction of cooperative biological sensors \cite{barato2014efficiency,
becker2015optimal}, but did not use the more general class of semi-Markov input
and did not calculate the full suite of metrics here, leaving much sensor
operation untouched.



An example---$\ms_l=0.5,~\ms_h=2.0$ and $k_O=k_C=1.0,~n=2$---illustrates
that roughly $99\%$ of $\Ifut$ is devoted to instantaneous memory $\Imem$ and
roughly $25\%$ of $\beta P$ is devoted to $\Inpdot$. That is, the inefficiency
in choosing what information to store about the present input contributes
greatly to energetic inefficiency. These results hold qualitatively even when
the dwell-time distributions are log-normal, i.e., are heavier-tailed.

Fig.~\ref{fig:hill} shows that increased cooperativity---that is, increases in
$n$---lead to increases in predictive performance, qualitatively in line with
Ref. \cite{martins2011trade}. Additionally, the larger the cooperativity, the
higher the fraction of $\Imem/\Ifut$. Larger cooperativity, however, leads to
roughly linear increases in the power consumption and the nonpredictive
information rate, whereas increases in predictive power take a more sigmoidal
shape. We therefore might prefer intermediate values of cooperativity (e.g.,
$n\approx 5$) to larger values of cooperativity (e.g., $n\geq 10$). This is
qualitatively similar to the results of Refs. \cite{tkavcik2009optimizing,
walczak2010optimizing}, in that physical constraints can force optimal
information transmission at intermediate levels of cooperativity.

\begin{figure}
\centering
\includegraphics[width=\columnwidth]{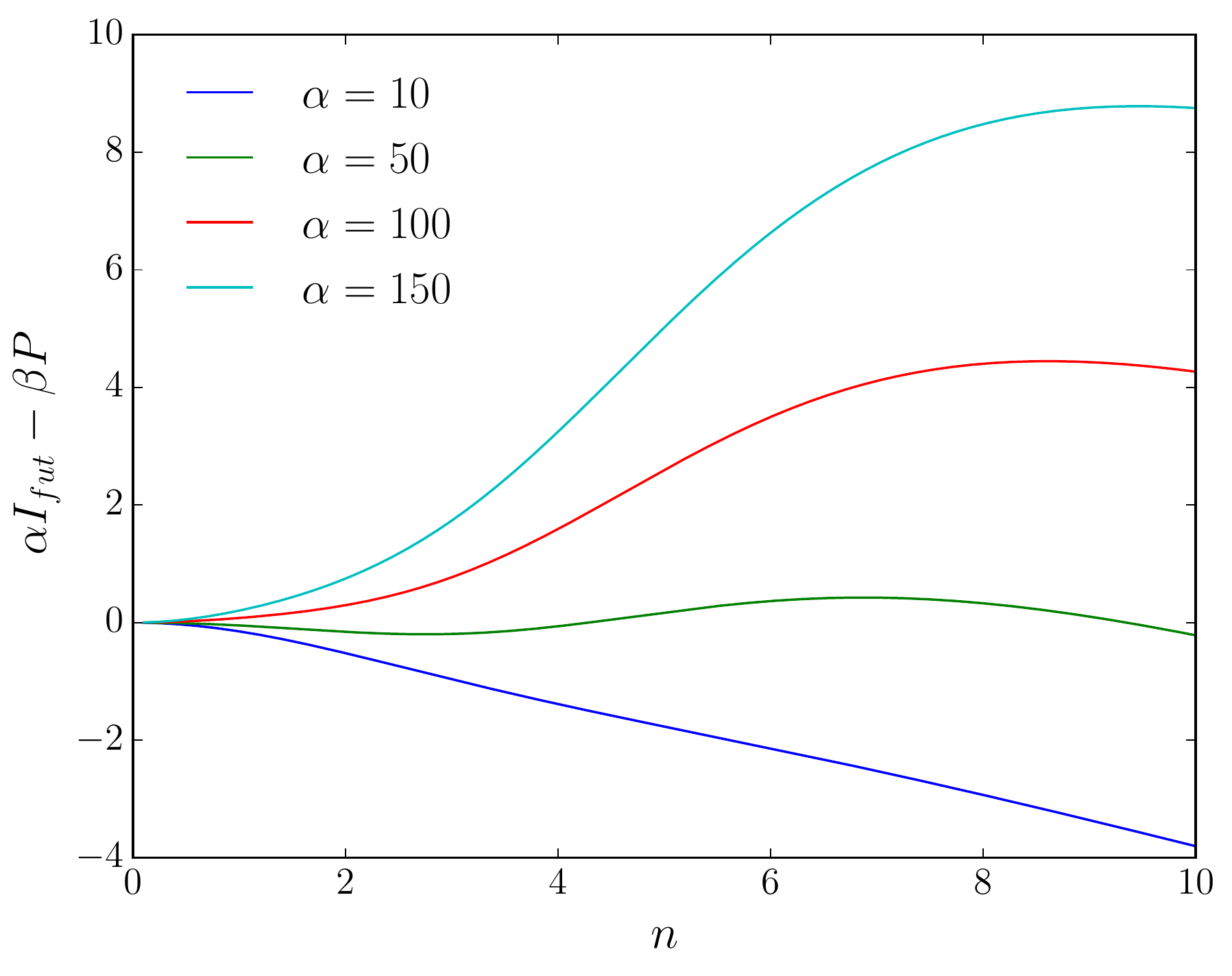}
\includegraphics[width=\columnwidth]{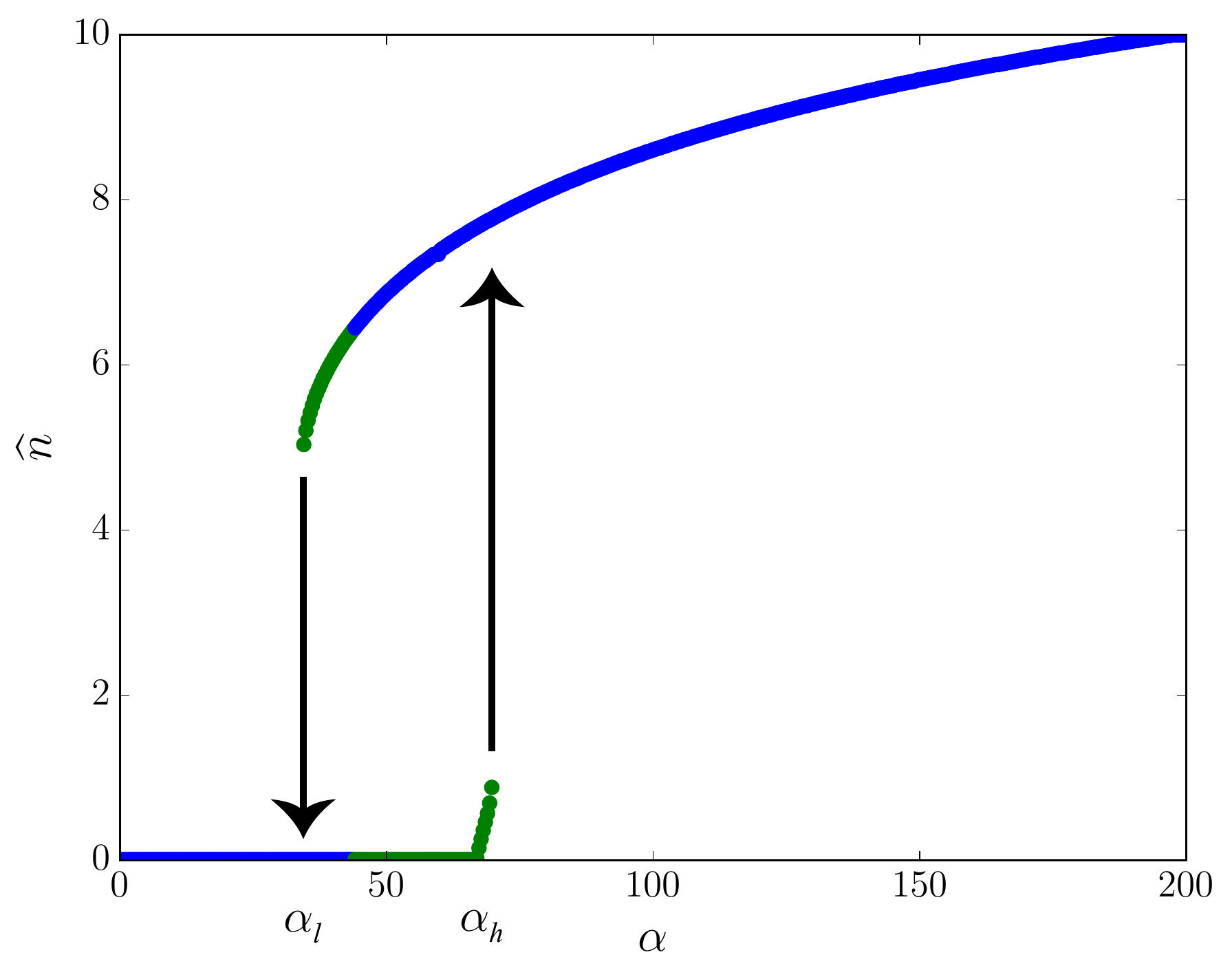}
\caption{(Top) ``Lagrangian'' $\alpha I_{fut} + \beta P$ for several arbitrary
	$\alpha$---that represent the energy rewards of $I_{fut}$---as a function of
	$n$. Notice that a particular $\alpha$ singles out a particular optimal
	$\widehat{n}$.
	(Bottom) Optimal cooperativity $\widehat{n}$ as a function of conversion
	factor $\alpha$.
	Circles (solid blue) are the values of the global maximum $\widehat{n}$
	and circles (solid green) are local maxima.
	Arrows indicate the directionality in the hysteresis loop:
	$\widehat{n}$ jumps up at the upper value $\alpha_h$, if starting at low
	$\alpha$, and jumps down at the lower value $\alpha_l$, if starting at high
	$\alpha$.
	}
\label{fig:Efficiency}
\end{figure}

In the absence of a reward function, we assert that energetic rewards are
proportional to $I_{fut}$ \cite{Stil07c, tishby2011information}, so that the
total energy budget is $\alpha \Ifut-\beta P$.  The proportionality constant
$\alpha$ is set by the type of environment in which one finds oneself. Then, the
total energy budget $\alpha I_{fut}-\beta P$ is optimized by the cooperativity
$\widehat{n} = \arg\max_n \left( \alpha I_{fut}-\beta P \right)$. As both
$I_{fut}$ and $\beta P$ increase monotonically with $n$, there is generically
only one such cooperativity $\widehat{n}$. At lower $\alpha$, though, there are
two local maxima of the function of $n$ given by $\alpha I_{fut}-\beta P$, as
shown at Fig. \ref{fig:Efficiency}(Top). Let's pursue the consequences of this
regime dependence.

There are rules for how sensor biochemical parameters adapt to
the present environment. If adaptation rules for cooperativity of a Hill
molecule increase the total energy budget by gradient descent then, for a range
of $\alpha$, we expect that the cooperativity of the Hill molecule to be in
either of the two local maxima of $\alpha \Ifut-\beta P$ just noted.  We assume a separation of timescales---namely, that cooperativity adapts much more slowly than the longest environmental timescale. Figure
\ref{fig:Efficiency}(Bottom) shows the optimal cooperativity $\widehat{n}$ as a
function of conversion factor $\alpha$. As expected from the presence of two
local maxima at lower $\alpha$, there is a discontinuity (supercritical
bifurcation \cite{Stro94a}) in the function of $\alpha$ given by $\arg\max_n
\alpha I_{fut}-\beta P$. Thus, initially, if $\alpha$ (energetic reward for
prediction) increases, the cooperativity discontinuously jumps to a higher
$\widehat{n}$ at a critical $\alpha_h$. From there, if one decreases $\alpha$,
optimal cooperativity slowly decreases, but stays high well below $\alpha_h$,
suddenly decreasing to zero cooperativity at the lower value $\alpha_l$. Thus,
there is a substantial hysteresis loop built into the optimal trade-off between
energy and sensitivity.

Recall that in switching circuits hysteresis is essential to adding stability
to a switch's response. Hysteresis stops ``race'' conditions in which the
switch oscillates wildly just as the threshold is passed, amplifying any noise
in the control and internal dynamics. In the Hill molecule, hysteresis is
helpful if a memory of past environmental conditions ($\alpha$) provides
insight into future conditions (future $\alpha$). For example, the environment
might shift $\alpha$ suddenly to being low, but there is a replenishment
mechanism for the available energy that will soon increase $\alpha$ again.
Thus, we see that robustness to environmental noise emerges naturally as the
sensor adapts to and anticipates changing external conditions.

\paragraph*{\textbf{Conclusion}}
We provided closed-form expressions for instantaneous memory, total predictable
information, nonpredictive information rate, and power consumption for a
conditionally Markovian channel subject to unifilar hidden semi-Markov input.

In motivating these metrics for prediction and dissipation on general
timescales, we appealed to an extension of Kelly's classic bet-hedging that
arises from his information-theoretic analysis of trading-off the benefits of
risky, but highly profitable resource investment against the costs of sudden
loss \cite{Cove06a, bergstrom2004shannon}. Here, a sensor faces an analogous
challenge of high sensitivity but at an energy cost that might be suddenly
wasted when the environmental conditions fluctuate. Similar bet-hedging
strategies have been implicated in other biological systems, such as in seed
germination in annual plants \cite{Bulm84a} and bacteriophages \cite{Masl15a}
and in population biology \cite{Cohe66a,Metz92a} and evolution
\cite{bergstrom2004shannon,Sege87a} more generally. The present setting,
though, implicates such strategy optimization in a substantially more
elementary and primitive biological subsystem.

Finally, we used these formulae to calculate the predictive performance and
energetic inefficiency of a simple model of a biological sensor---a Hill
molecule. We found that increases in cooperativity yield increases in both
predictive performance and energy consumption and that the relative balance
between those increases naturally leads to sensor robustness to environmental
fluctuations supported by dynamical hysteresis. Given the Hill molecule's
simplicity as a model sensor, we expect to find hysteresis and the resulting
robustness in more complex biological sensory systems.

The ease with which these various metrics were calculated masks the difficulty
of obtaining the necessary closed-form expressions. (Cf. Supplementary
Materials.) We provided universal estimators for various predictive and
dissipative metrics for conditionally Markovian channels, as unifilar hidden
semi-Markov processes are that general. One practical consequence it that those
wishing to study the relationship between prediction and dissipation need not
simulate arbitrarily long trajectories. Instead, they can validate or
invalidate predictive learning rules and sensor designs using the universal
estimators of these predictive and dissipative metrics. Then they can
efficiently search through parameter space for ``optimal'' (predictive and
energy-efficient) sensors. In addition, given that the theories of random
dynamical systems and of input-dependent dynamical systems are still under
development \cite{arnold2013random}, we believe the formulae presented here
will lead in those domains to a precise generalization of time-scale matching
for nonlinear systems \cite{becker2015optimal}.


\acknowledgments

We thank N. Ay, A. Bell, W. Bialek, S. Dedeo, C. Hillar, I. Nemenman, and S.
Still for useful conversations and the Santa Fe Institute for its hospitality
during visits, where JPC is an External Faculty member. This material is based
upon work supported by, or in part by, the John Templeton Foundation grant
52095, the Foundational Questions Institute grant FQXi-RFP-1609, the U. S.
Army Research Laboratory and the U. S. Army Research Office under contract
W911NF-13-1-0390. S.E.M. was funded by an MIT Physics of Living Systems
Fellowship.


\bibliography{chaos}

\begin{thebibliography}{10}

\bibitem{bennett1982thermodynamics}
C.~H. Bennett.
\newblock The thermodynamics of computation: a review.
\newblock {\em Intl. J. Theo. Physics}, 21(12):905--940, 1982.

\bibitem{Maxw88a}
J.~C. Maxwell.
\newblock {\em Theory of Heat}.
\newblock Longmans, Green and Co., London, United Kingdom, ninth edition, 1888.

\bibitem{Szil29a}
L.~Szilard.
\newblock On the decrease of entropy in a thermodynamic system by the
  intervention of intelligent beings.
\newblock {\em Z. Phys.}, 53:840--856, 1929.

\bibitem{mandal2012work}
D.~Mandal and C.~Jarzynski.
\newblock Work and information processing in a solvable model of {Maxwell's}
  demon.
\newblock {\em Proc. Natl. Acad. Sci. USA}, 109(29):11641--11645, 2012.

\bibitem{deffner2013information}
S.~Deffner and C.~Jarzynski.
\newblock Information processing and the second law of thermodynamics: An
  inclusive, hamiltonian approach.
\newblock {\em Phys. Rev. X}, 3(4):041003, 2013.

\bibitem{Note1}
Here, when analyzing sensory information processing in biological systems, we
  take care to distinguish intrinsic, functional, and useful computation \cite
  {Crut88a,Crut92c,Crutchfield&Mitchell94a}. \protect \emph {Intrinsic
  computation} refers to how a physical system stores and transforms its
  historical information. We take \protect \emph {functional computation} as
  information processing in a physical device that promotes the performance of
  a larger, encompassing system. Whereas, we take \protect \emph {useful
  computation} as information processing in a physical device used to achieve
  an external user's goal. The first is well-suited to analyzing structure in
  physical processes and determining if they are candidate substrates for any
  kind of information processing. The second is well-suited for discussing
  biological sensors, while the third is well-suited for discussing the
  benefits of contemporary digital computers.

\bibitem{parrondo2015thermodynamics}
J.~M.~R. Parrondo, J.~M. Horowitz, and T.~Sagawa.
\newblock Thermodynamics of information.
\newblock {\em Nat. Physics}, 11(2):131--139, 2015.

\bibitem{Agha16d}
C.~Aghamohammdi and J.~P. Crutchfield.
\newblock Thermodynamics of random number generation.
\newblock {\em Phys. Rev. E}, 95(6):062139, 2017.

\bibitem{hinczewski2014cellular}
M.~Hinczewski and D.~Thirumalai.
\newblock Cellular signaling networks function as generalized
  {Wiener-Kolmogorov} filters to suppress noise.
\newblock {\em Phys. Rev. X}, 4(4):041017, 2014.

\bibitem{Boyd16c}
A.~B. Boyd, D.~Mandal, and J.~P. Crutchfield.
\newblock Correlation-powered information engines and the thermodynamics of
  self-correction.
\newblock {\em Phys. Rev. E}, 95(1):012152, 2017.

\bibitem{Still2012}
S.~Still, D.~A. Sivak, A.~J. Bell, and G.~E. Crooks.
\newblock Thermodynamics of prediction.
\newblock {\em Phys. Rev. Lett.}, 109:120604, Sep 2012.

\bibitem{barato2014efficiency}
A.~C. Barato, D.~Hartich, and U.~Seifert.
\newblock Efficiency of cellular information processing.
\newblock {\em New J. Physics}, 16(10):103024, 2014.

\bibitem{Horowitz2014}
J.~M. Horowitz and M.~Esposito.
\newblock Thermodynamics with continuous information flow.
\newblock {\em Phys. Rev. X}, 4:031015, Jul 2014.

\bibitem{Boyd16d}
A.~B. Boyd, D.~Mandal, and J.~P. Crutchfield.
\newblock Leveraging environmental correlations: The thermodynamics of
  requisite variety.
\newblock {\em J. Stat. Phys.}, 167(6):1555--1585, 2016.

\bibitem{goldt2017stochastic}
S.~Goldt and U.~Seifert.
\newblock Stochastic thermodynamics of learning.
\newblock {\em Phys. Rev. Lett.}, 118(1):010601, 2017.

\bibitem{Boyd16e}
A.~B. Boyd, D.~Mandal, P.~M. Riechers, and J.~P. Crutchfield.
\newblock Transient dissipation and structural costs of physical information
  transduction.
\newblock {\em Phys. Rev. Lett.}, 118:220602, 2017.

\bibitem{Boyd14b}
A.~B. Boyd and J.~P. Crutchfield.
\newblock Maxwell demon dynamics: {Deterministic} chaos, the {Szilard} map, and
  the intelligence of thermodynamic systems.
\newblock {\em Phys. Rev. Lett.}, 116:190601, 2016.

\bibitem{rao1999predictive}
R.~P.~N. Rao and D.~H. Ballard.
\newblock Predictive coding in the visual cortex: a functional interpretation
  of some extra-classical receptive-field effects.
\newblock {\em Nat. Neurosci.}, 2(1):79--87, 1999.

\bibitem{Palm13a}
S.~E. Palmer, O.~Marre, M.~J. Berry, and W.~Bialek.
\newblock Predictive information in a sensory population.
\newblock {\em Proc. Natl. Acad. Sci. USA}, 112(22):6908--6913, 2015.

\bibitem{chklovskii2004maps}
D.~B. Chklovskii and A.~A. Koulakov.
\newblock Maps in the brain: what can we learn from them?
\newblock {\em Annu. Rev. Neurosci.}, 27:369--392, 2004.

\bibitem{hasenstaub2010metabolic}
A.~Hasenstaub, S.~Otte, E.~Callaway, and T.~J. Sejnowski.
\newblock Metabolic cost as a unifying principle governing neuronal biophysics.
\newblock {\em Proc. Natl. Acad. Sci. USA}, 107(27):12329--12334, 2010.

\bibitem{Note2}
We only consider online or real-time computations, so that the oft-considered
  energy-speed-accuracy tradeoff \cite {lan2012energy, lahiri2016universal}
  reduces to an energy-accuracy tradeoff.

\bibitem{sutton1998reinforcement}
R.~S. Sutton and A.~G. Barto.
\newblock {\em Reinforcement learning: An introduction}.
\newblock MIT Press, Cambridge, Massachusetts, 1998.

\bibitem{littman2001predictive}
M.~L. Littman, R.~S. Sutton, and S.~P. Singh.
\newblock In {\em NIPS}, volume~14, pages 1555--1561, 2001.

\bibitem{Brodu11}
N.~Brodu.
\newblock {\em Adv. Complex Sys.}, 14(05):761--794, 2011.

\bibitem{little2014learning}
D.~Y. Little and F.~T. Sommer.
\newblock Learning and exploration in action-perception loops.
\newblock {\em Closing the Loop Around Neural Systems}, page 295, 2014.

\bibitem{becker2015optimal}
N.~B. Becker, A.~Mugler, and P.~R. ten Wolde.
\newblock Optimal prediction by cellular signaling networks.
\newblock {\em Phys. Rev. lett.}, 115(25):258103, 2015.

\bibitem{creutzig2008predictive}
F.~Creutzig and H.~Sprekeler.
\newblock Predictive coding and the slowness principle: An
  information-theoretic approach.
\newblock {\em Neural Comp.}, 20(4):1026--1041, 2008.

\bibitem{creutzig2009past}
F.~Creutzig, A.~Globerson, and N.~Tishby.
\newblock Past-future information bottleneck in dynamical systems.
\newblock {\em Phys. Rev. E}, 79(4):041925, 2009.

\bibitem{Marz17b}
S.~Marzen and J.~P. Crutchfield.
\newblock Structure and randomness of continuous-time discrete-event processes.
\newblock 2016.
\newblock arxiv.org:1704.04707.

\bibitem{izhikevich2007dynamical}
E.~M. Izhikevich.
\newblock {\em Dynamical systems in neuroscience}.
\newblock MIT press, Cambridge, Massachusetts, 2007.

\bibitem{Shal98a}
C.~R. Shalizi and J.~P. Crutchfield.
\newblock Computational mechanics: Pattern and prediction, structure and
  simplicity.
\newblock {\em J. Stat. Phys.}, 104:817--879, 2001.

\bibitem{Jame11a}
R.~G. James, C.~J. Ellison, and J.~P. Crutchfield.
\newblock Anatomy of a bit: {Information} in a time series observation.
\newblock {\em CHAOS}, 21(3):037109, 2011.

\bibitem{Stil07b}
S.~Still, J.~P. Crutchfield, and C.~J. Ellison.
\newblock Optimal causal inference: {Estimating} stored information and
  approximating causal architecture.
\newblock {\em CHAOS}, 20(3):037111, 2010.

\bibitem{Yeun91a}
R.~W. Yeung.
\newblock {A New Outlook on Shannon's Information Measures}.
\newblock {\em IEEE Trans. Info. Th.}, 37(3):466--474, 1991.

\bibitem{white2004short}
O.~L. White, D.~D. Lee, and H.~Sompolinsky.
\newblock Short-term memory in orthogonal neural networks.
\newblock {\em Phys. Rev. Lett.}, 92(14):148102, 2004.

\bibitem{Cove06a}
T.~M. Cover and J.~A. Thomas.
\newblock {\em Elements of Information Theory}.
\newblock Wiley-Interscience, New York, second edition, 2006.

\bibitem{schnakenberg1976network}
J~Schnakenberg.
\newblock Network theory of microscopic and macroscopic behavior of master
  equation systems.
\newblock {\em Rev. Mod. Physics}, 48(4):571, 1976.

\bibitem{marzen2013statistical}
S.~Marzen, H.~G. Garcia, and R.~Phillips.
\newblock Statistical mechanics of monod--wyman--changeux (mwc) models.
\newblock {\em J. Mole. Bio.}, 425(9):1433--1460, 2013.

\bibitem{tkavcik2009optimizing}
G.~Tka{\v{c}}ik, A.~M. Walczak, and W.~Bialek.
\newblock Optimizing information flow in small genetic networks.
\newblock {\em Phys. Rev. E}, 80(3):031920, 2009.

\bibitem{walczak2010optimizing}
A.~M. Walczak, G.~Tka{\v{c}}ik, and W.~Bialek.
\newblock Optimizing information flow in small genetic networks. {II.}
  {Feed-forward} interactions.
\newblock {\em Phys. Rev. E}, 81(4):041905, 2010.

\bibitem{martins2011trade}
B.~M.~C. Martins and P.~S. Swain.
\newblock Trade-offs and constraints in allosteric sensing.
\newblock {\em PLoS Comput. Bio.}, 7(11):e1002261, 2011.

\bibitem{Stil07c}
S.~Still.
\newblock Information-theoretic approach to interactive learning.
\newblock {\em EuroPhys. Lett.}, 85:28005, 2009.
\newblock arxiv.org physics 0709.1948.

\bibitem{tishby2011information}
Naftali Tishby and Daniel Polani.
\newblock Information theory of decisions and actions.
\newblock In {\em Perception-action cycle}, pages 601--636. Springer, 2011.

\bibitem{Stro94a}
S.~H. Strogatz.
\newblock {\em Nonlinear Dynamics and Chaos: with applications to physics,
  biology, chemistry, and engineering}.
\newblock Addison-Wesley, Reading, Massachusetts, 1994.

\bibitem{bergstrom2004shannon}
C.~T. Bergstrom and M.~Lachmann.
\newblock Shannon information and biological fitness.
\newblock In {\em Information Theory Workshop, 2004. IEEE}, volume IEEE
  0-7803-8720-1, pages 50--54. IEEE, 2004.

\bibitem{Bulm84a}
M.~G. Bulmer.
\newblock Delayed germination of seeds: {Cohen's} model revisited.
\newblock {\em Theo. Pop. Bio.}, 26:367--377, 1984.

\bibitem{Masl15a}
S.~Maslov and K.~Sneppen.
\newblock Well-temperate phage: optimal bet-hedging against local environmental
  collapses.
\newblock {\em Sci. Reports}, 5:10523, 2015.

\bibitem{Cohe66a}
D.~Cohen.
\newblock Optimizing reproduction in a randomly varying environment.
\newblock {\em J. Theo. Bio.}, 12:119--129, 1966.

\bibitem{Metz92a}
J.~A.~J. Metz, R.~M. Nisbet, and S.~A.~H. Geritz.
\newblock How should we define fitness for general ecological scenarios?
\newblock {\em Trends Ecol. Evol.}, 7:198--202, 1992.

\bibitem{Sege87a}
J.~Seger and H.~J. Brockmann.
\newblock What is bet-hedging?
\newblock {\em Oxford Surveys in Evolutionary Biology}, 4:182--211, 1987.

\bibitem{arnold2013random}
L.~Arnold.
\newblock {\em Random dynamical systems}.
\newblock Springer Science \& Business Media, New York, New York, 2013.

\bibitem{Crut88a}
J.~P. Crutchfield and K.~Young.
\newblock Inferring statistical complexity.
\newblock {\em Phys. Rev. Let.}, 63:105--108, 1989.

\bibitem{Crut92c}
J.~P. Crutchfield.
\newblock The calculi of emergence: Computation, dynamics, and induction.
\newblock {\em Physica D}, 75:11--54, 1994.

\bibitem{Crutchfield&Mitchell94a}
J.~P. Crutchfield and M.~Mitchell.
\newblock The evolution of emergent computation.
\newblock {\em Proc. Natl. Acad. Sci.}, 92:10742--10746, 1995.

\bibitem{lan2012energy}
G.~Lan, P.~Sartori, S.~Neumann, V.~Sourjik, and Y.~Tu.
\newblock The energy-speed-accuracy trade-off in sensory adaptation.
\newblock {\em Nat. Physics}, 8(5):422--428, 2012.

\bibitem{lahiri2016universal}
S.~Lahiri, J.~Sohl-Dickstein, and S.~Ganguli.
\newblock A universal tradeoff between power, precision and speed in physical
  communication.
\newblock {\em arXiv:1603.07758}, 2016.

\end{thebibliography}

\clearpage
\begin{center}
\large{Supplementary Materials}\\
\emph{Prediction and Power in Molecular Sensors}\\
Sarah E. Marzen and James P. Crutchfield
\end{center}

\setcounter{equation}{0}
\setcounter{page}{1}
\makeatletter
\renewcommand{\theequation}{S\arabic{equation}}

\section{Extending Kelly's argument}

Although the use of mutual information for a biological sensor may seem
arbitrary, it gains operational significance via a straightforward extension of
Kelly's bet-hedging arguments \cite[Ch. 6 of]{Cove06a}. Here, we switch to a
discrete-time analysis. Kelly's classic result states that a population of
organisms increases its expected log growth rate by $I[\Rep_t;\MS_{t+1}]$---the
\emph{instantaneous predictive information}. Each organism stores information
about the past environments in a sensory variable $\rep$ and chooses
stochastically from $p(g|\rep)$ to exhibit phenotype $g$ based on this sensory
variable. Kelly's original derivation assumed that only one phenotype can
reproduce in each possible environment. We extend this result by relaxing this
assumption, following Ref. \cite{bergstrom2004shannon}. Let $n_t$ be the number
of organisms at time $t$; let $p(g|\rep_t)$ be the probability that an organism
expresses phenotype $g$ given sensory state $\rep_t$; let $\ms_t$ be the
sensory input at time $t$; and let $f(g,\ms)$ be the growth rate of phenotype
$g$ in environment $\ms$. Then, we straightforwardly obtain:
\begin{align*}
n_{t+1} & = \sum_{g} (p(g|\rep_t)n_t) f(g,\ms_{t+1}) \\
        & = \left(\sum_g p(g|\rep_t) f(g,\ms_{t+1}) \right) n_t
  ~.
\end{align*}
This yields an expected log growth rate of:
\begin{align*}
r & = \left\langle \log\frac{n_{t+1}}{n_t} \right\rangle \\
  & = \left\langle \log \left(\sum_g p(g|\rep_t) f(g,\ms_{t+1}) \right)
  	\right\rangle \\
  & = \sum_{\rep_t,\ms_{t+1}} p(\rep_t,\ms_{t+1})
  \log \left(\sum_g p(g|\rep_t) f(g,\ms_{t+1}) \right)
  ~.
\end{align*}
We seek the bet-hedging strategy that maximizes expected log growth rate. That
is, maximize $r$, subject to the constraint that $\sum_g p(g|\rep_t) = 1$ for
all $\rep_t$, via the Lagrangian:
\begin{align*}
\mathcal{L} & = \sum_{\rep_t,\ms_{t+1}} p(\rep_t,\ms_{t+1})
  \log \left(\sum_g p(g|\rep_t) f(g,\ms_{t+1}) \right) \nonumber \\
  & \qquad + \sum_{\rep_t} \lambda_{\rep_t} \sum_g p(g|\rep_t)
  ~,
\end{align*}
with respect to $p(g|\rep_t)$. Following Ref. \cite{bergstrom2004shannon}, let
$\bf{x}$ be the vector of optimal $p(g|\rep_t)$, let $\bf{p}$ be the vector of
$p(\ms|\rep_t)$, and let $W$ be the matrix with elements $f(g,\ms)$. Then, we
find that:
\begin{align*}
(Wx)_k &= p_k/\sum_{j} (W^{-1})_{jk} \\
    \rightarrow \bf{x} & = W^{-1}
	\left(\bf{p} \odot [1^{\top}W^{-1}]^{\odot -1}\right)
\end{align*}
is the maximizing conditional distribution \emph{if} it is in the interior of
the simplex. This gives an expected log growth rate:
\begin{align*}
r^* & = \sum_{\rep_t,\ms_{t+1}}
  p(\rep_t,\ms_{t+1})
  \log \frac{p(\ms_{t+1}|\rep_t)}{\sum_{g} (W^{-1})_{g,\ms_{t+1}}} \\
  & = -H[\MS_{t+1}|\Rep_t]
  - \sum_{\ms_{t+1}} p(\ms_{t+1}) \log \sum_g ((W^{-1})_{g,\ms_{t+1}})
  ~.
\end{align*}
The difference between this expected log growth rate and the maximal expected
log growth rate of a population without any sensing capabilities is:
\begin{eqnarray}
\Delta r^*
  & = -H[\MS_{t+1}|\Rep_t] + H[\MS_{t+1}] \nonumber \\
  & = I[\Rep_t;\MS_{t+1}]
  ~.
\end{eqnarray}
This is exactly the instantaneous predictive information, which lower bounds the total predictable information calculated here.

\section{Revisiting the ``Thermodynamics of Prediction''}

For completeness, we review the derivation of Eq.~(\ref{eq:ToP}). Let
$\ms_t$ represent the input at time $t$, let $\rep_t$ represent the sensor
state at time $t$, and let $E(\ms,\rep)$ denote the system's energy function.
We assume constant temperature.  The system's temperature-normalized
nonequilibrium free energy $F_{neq}$ is given by:
\begin{align}
\beta F_{neq}[p(\ms,\rep)] = \beta \langle E(\ms,\rep)\rangle - H[\Rep|\MS]
  ~.
\label{eq:noneq}
\end{align}
The validity of Ref. \cite{Still2012}'s derivation rests on the nonequilibrium
free energy being a Lyapunov function. Intuitively, this corresponds to an
assumption that the system reduces its nonequilibrium free energy when the
sensor thermalizes. If so, then:
\begin{align*}
\beta F_{neq}[p(\ms_{t+\Delta t},\rep_t)]
  \geq \beta F_{neq}[p(\ms_{t+\Delta t},\rep_{t+\Delta t})]
  ~,
\end{align*}
giving:
\begin{align*}
0 & \leq \beta F_{neq}[p(\ms_{t+\Delta t},\rep_t)]
  -\beta F_{neq}[p(\ms_{t+\Delta t},\rep_{t+\Delta t})] \\
  & \leq \left(\beta \langle E(\ms_{t+\Delta t},\rep_t)\rangle
  - \H[\Rep_t|\MS_{t+\Delta t}]\right) \nonumber \\
  & \qquad - \left(\beta \langle E(\ms_{t+\Delta t},\rep_{t+\Delta t})\rangle
  - \H[\Rep_{t+\Delta t}|\MS_{t+\Delta t}]\right)
\end{align*}
and, from stationarity, $\langle E(\ms_{t+\Delta t},\rep_{t+\Delta
t})\rangle=\langle E(\ms_{t},\rep_{t})\rangle$ and $H[\Rep_{t+\Delta
t}|\MS_{t+\Delta t}] = H[\Rep_{t}|\MS_{t}]$, giving:
\begin{align*}
0 & \leq \beta \left(\langle E(\ms_{t+\Delta t},\rep_t)\rangle
  - \langle E(\ms_t,\rep_t)\rangle\right) \nonumber \\
  & \qquad - \left(H[\Rep_t|\MS_{t+\Delta t}]-H[\Rep_t|\MS_t]\right) \\
  & \leq \lim_{\Delta t\rightarrow 0}
  \frac{\beta \left(\langle E(\ms_{t+\Delta t},\rep_t)\rangle - \langle E(\ms_t,\rep_t)\rangle\right)}{\Delta t} \nonumber \\
  & \qquad - \lim_{\Delta t\rightarrow 0} \frac{H[\Rep_t|\MS_{t+\Delta t}]-H[\Rep_t|\MS_t]}{\Delta t}
  ~.
\end{align*}
We recognize the first term as the temperature-normalized power $\beta P$.
Hence, the nonpredictive information rate is the increase in unpredictability
of sensor state $\Rep_t$ given a slightly delayed environmental state:
\begin{align}
\Inpdot := \lim_{\Delta t\rightarrow 0}
  \frac{\H[\Rep_t|\MS_{t+\Delta t}]-\H[\Rep_t|\MS_t]}{\Delta t} \leq \beta P
  ~.
\label{eq:top1}
\end{align}
From standard information theory identities \cite{Cove06a}---namely, $\I[U;V]=
\H[U]-\H[U|V]$---we see that:
\begin{align*}
\H[\Rep_t|\MS_{t+\Delta t}] - \H[\Rep_t|\MS_t]
  = \I[\Rep_t;\MS_t] - \I[\Rep_t;\MS_{t+\Delta t}]
  ~.
\end{align*}
Reference \cite{Still2012}'s main result  follows directly:
\begin{align}
\Inpdot = \lim_{\Delta t\rightarrow 0}
  \frac{\I[\Rep_t;\MS_t] - \I[\Rep_t;\MS_{t+\Delta t}]}{\Delta t}
  \leq \beta P
  ~.
\label{eq:top}
\end{align}

Differences in presentation come from the difference between discrete- and
continuous-time formulations. To make this clear, we present a continuous-time
formulation of the same result, following Ref. \cite{Horowitz2014}. We start
from $\beta F_{neq}[p(x_t,y_{t'})]$ being a Lyapunov function in $t'$:
\begin{align*}
0 & \geq \beta \frac{\partial F_{neq}[p(x_t,y_{t'})]}{\partial t'} \\
  & = \beta \frac{\partial}{\partial t'}
  \left\langle E(x_t,y_{t'})\right\rangle\Big|_{t'=t}
  - \frac{\partial}{\partial t'} \H[Y_{t'}|X_t]|_{t'=t} \\
  & = \beta \left\langle \frac{\partial E(x_t,y_t)}{\partial y_t}
  \dot{y}_t \right\rangle
  -\frac{\partial}{\partial t'} \H[Y_{t'}|X_t]|_{t'=t}
  ~.
\end{align*}
We then recognize $\beta \left\langle \frac{\partial E(x_t,y_t)}{\partial y_t}
\dot{y}_t\right\rangle$ as the temperature-normalized rate of heat dissipation
$\beta\dot{Q}$, so that:
\begin{align*}
\beta \dot{Q} \leq \frac{\partial}{\partial t'} \H[Y_{t'}|X_t]|_{t'=t}
  ~.
\end{align*}
In nonequilibrium steady state, $\frac{d}{dt}\langle E\rangle = 0$ and
$\frac{d}{dt} \H[Y_t|X_t] = 0$. As a result, $\beta\dot{Q}+\beta P = 0$ and:
\begin{align*}
\frac{\partial}{\partial t'} \H[Y_{t'}|X_t]|_{t'=t}
  = -\frac{\partial}{\partial t'} \H[Y_t|X_{t'}]|_{t'=t}
  ~,
\end{align*}
giving:
\begin{align}
\beta P \geq \frac{\partial}{\partial t'} \H[Y_t|X_{t'}]|_{t'=t}
  ~,
\label{eq:top1CT}
\end{align}
which we recognize as the continuous-time formulation of Eq.~(\ref{eq:top1}).
Again invoking stationarity, $\frac{d}{dt} \H[X_t] = 0$, and so:
\begin{align}
\beta P \geq -\frac{\partial}{\partial t'} \I[X_{t'};Y_t]|_{t'=t}
  ~,
\label{eq:topCT}
\end{align}
the continuous-time formulation of Eq.~(\ref{eq:top}). We have, in
Eqs.~(\ref{eq:top1}), (\ref{eq:top}), (\ref{eq:top1CT}), and (\ref{eq:topCT}),
four equivalent definitions for the nonpredictive information rate in the
nonequilibrium steady state limit.

\begin{widetext}

\section{Closed-form Expressions for Unifilar Hidden Semi-Markov
Environments}

To find $\rho(\fst,\rep)$, we start with the following:
\begin{align}
\Prob(\FSt_{t+\Delta t} & =(g,\ms,\tau),\Rep_{t+\Delta t}=\rep)
  \nonumber \\
  & = \sum_{g',\ms,',\tau',\rep'}
  \Prob(\FSt_{t+\Delta t}=(g,\ms,\tau),
    \Rep_{t+\Delta t}=\rep|\FSt_t=(g',\ms',\tau'),\Rep_t=\rep')
  \Prob(\FSt_t=(g',\ms',\tau'),\Rep_t=\rep')
  ~.
\label{eq:main}
\end{align}
We decompose the transition probability using the feedforward nature of the
transducer as:
\begin{align*}
\Prob(\FSt_{t+\Delta t}=(g,\ms,\tau), \Rep_{t+\Delta t}=\rep
  |\FSt_t=(g',\ms',\tau'),\Rep_t=\rep')
  & = \Prob(\FSt_{t+\Delta t}=(g,\ms,\tau)|\FSt_t=(g',\ms',\tau')) \nonumber\\
  & \qquad \times\Prob(\Rep_{t+\Delta t}=\rep|\FSt_t=(g',\ms',\tau'),\Rep_t=\rep')
  ~.
\end{align*}
From the setup, we have:
\begin{align*}
\Prob(\Rep_{t+\Delta t}=\rep|\FSt_t=(g',\ms',\tau'),\Rep_t=\rep')
  = \begin{cases}
  k_{\rep'\rightarrow\rep}(\ms')\Delta t & \rep\neq\rep' \\
  1-k_{\rep'\rightarrow\rep'}(\ms')\Delta t & \rep=\rep'
  \end{cases}
  ~,
\end{align*}
with corrections of $O(\Delta t^2)$.

Now split this into two cases. As long as $\tau > \Delta t$, so that
$\ms=\ms'$, we have:
\begin{align*}
\Prob(\FSt_{t+\Delta t}=(g,\ms,\tau)|\FSt_t=(g',\ms',\tau')) = \frac{\Phi_{g}(\tau)}{\Phi_{g}(\tau')} \delta (\tau-(\tau'+\Delta t)) \delta_{\ms,\ms'} \delta_{g,g'}
  ~.
\end{align*}
Then, Eq.~(\ref{eq:main}) reduces to
\begin{align}
\Prob(\FSt_{t+\Delta t}=(g,\ms,\tau),\Rep_{t+\Delta t}=\rep)
  & = \sum_{\rep'} \Prob(\FSt_{t+\Delta t}=(g,\ms,\tau)|\FSt_t=(g,\ms,\tau-\Delta t)) \Prob(\Rep_{t+\Delta t}=\rep|\FSt_t=(g,\ms,\tau-\Delta t),\Rep_t=\rep')
  \nonumber \\
  & \qquad \times \Prob(\FSt_t=(g,\ms,\tau-\Delta t),\Rep_t=\rep') \nonumber \\
  & = \sum_{\rep'\neq \rep}\Prob(\FSt_{t+\Delta t}=(g,\ms,\tau)|\FSt_t=(g,\ms,\tau-\Delta t)) \Prob(\Rep_{t+\Delta t}=\rep|\FSt_t=(g,\ms,\tau-\Delta t),\Rep_t=\rep')
  \nonumber \\
  & \qquad \times \Prob(\FSt_t=(g,\ms,\tau-\Delta t),\Rep_t=\rep') 
  \nonumber \\
  & \qquad + \Prob(\FSt_{t+\Delta t}=(g,\ms,\tau)|\FSt_t=(g,\ms,\tau-\Delta t)) \Prob(\Rep_{t+\Delta t}=\rep|\FSt_t=(g,\ms,\tau-\Delta t),\Rep_t=\rep) 
  \nonumber \\
  & \qquad \times \Prob(\FSt_t=(g,\ms,\tau-\Delta t),\Rep_t=\rep)
  \nonumber \\
  & = \sum_{\rep'\neq\rep} \frac{\Phi_{g}(\tau)}{\Phi_{g}(\tau-\Delta t)} k_{\rep'\rightarrow\rep}(\ms) \Prob(\FSt_t=(g,\ms,\tau-\Delta t),\Rep_t=\rep') \Delta t \nonumber \\
  &\qquad + \frac{\Phi_{g}(\tau)}{\Phi_{g}(\tau-\Delta t)} \left(1-k_{\rep\rightarrow\rep}(\ms)\Delta t\right) \Prob(\FSt_t=(g,\ms,\tau-\Delta t),\Rep_t=\rep)
  ~,
\label{eq:main2}
\end{align}
plus terms of $O(\Delta t^2)$.  We Taylor expand $\Phi_{g}(\tau+\Delta t) =
\Phi_{g}(\tau) - \phi_{g}(\tau)\Delta t$ to find:
\begin{align*}
\frac{\Phi_{g}(\tau)}{\Phi_{g}(\tau-\Delta t)}
  = 1-\frac{\phi_{g}(\tau)}{\Phi_{g}(\tau)}\Delta t
  `,
\end{align*}
plus terms of $O(\Delta t^2)$. And, similarly, assuming differentiability, we
write:
\begin{align*}
\Prob(\FSt_t=(g,\ms,\tau-\Delta t),\Rep_t=\rep') = \Prob(\FSt_t=(g,\ms,\tau),\Rep_t=\rep') - \frac{d}{d\tau} \Prob(\FSt_t=(g,\ms,\tau),\Rep_t=\rep') \Delta t
  `,
\end{align*}
plus terms of $O(\Delta t^2)$.
Substitution into Eq.~(\ref{eq:main2}) then gives:
\begin{align*}
\Prob(\FSt_{t+\Delta t}=(g,\ms,\tau),\Rep_{t+\Delta t}=\rep)
  & = \left(\sum_{\rep'\neq\rep} k_{\rep'\rightarrow\rep}(\ms)
  \Prob(\FSt_t=(g,\ms,\tau),\Rep_t=\rep') \right) \Delta t
  + \Prob(\FSt_t=(g,\ms,\tau),\Rep_t=\rep) \\
  & \qquad - \frac{d\Prob(\FSt_t=(g,\ms,\tau),\Rep_t=\rep)}{d\tau}\Delta t
  - \frac{\phi_{g}(\tau)}{\Phi_{g}(\tau)}
  \Prob(\FSt_t=(g,\ms,\tau),\Rep_t=\rep) \Delta t \\
  & \qquad - k_{\rep\rightarrow\rep}(\ms)\Prob(\FSt_t=(g,\ms,\tau),\Rep_t=\rep) \Delta t
  ~,
\end{align*}
plus terms of $O(\Delta t^2)$. For notational ease, we denote:
\begin{align*}
\rho((g,\ms,\tau),\rep) := \Prob(\FSt_t=(\ms,\tau),\Rep_t=\rep)
  ~,
\end{align*}
which is equal to $\Prob(\FSt_{t+\Delta t}=(g,\ms,\tau),\Rep_{t+\Delta
t}=\rep)$ since we assumed the system is in a NESS. Then we have:
\begin{align*}
\rho((g,\ms,\tau),\rep)
  & = \left( \sum_{\rep'\neq \rep} k_{\rep'\rightarrow\rep}(\ms)
  \rho((g,\ms,\tau),\rep') \right) \Delta t + \rho((g,\ms,\tau),\rep) - \frac{d
  \rho((g,\ms,\tau),\rep)}{d\tau}\Delta t \\
  & \qquad - \frac{\phi_{g}(\tau)}{\Phi_g(\tau)} \rho((g,\ms,\tau),\rep) \Delta t 
  - k_{\rep\rightarrow\rep}(\ms) \rho((g,\ms,\tau),\rep)\Delta t
\end{align*}
plus corrections of $O(\Delta t^2)$.  We are left equating the coefficient of the $O(\Delta t)$ term to $0$:
\begin{align}
\frac{d\rho((g,\ms,\tau),\rep)}{d \tau}
  & = \sum_{\rep'\neq \rep} k_{\rep'\rightarrow \rep}(\ms) \rho((g,\ms,\tau),\rep')
  - \frac{\phi_g(\tau)}{\Phi_{g}(\tau)} \rho((g,\ms,\tau),\rep)
  - k_{\rep\rightarrow\rep}(\ms) \rho((g,\ms,\tau),\rep)
  ~.
\label{eq:diffyQ1}
\end{align}
Our task is simplified if we separate:
\begin{align*}
\rho((g,\ms,\tau),\rep) = p(\rep|g,\ms,\tau) \rho(g,\ms,\tau)
\end{align*}
and if we recall that;
\begin{align*}
\rho(g,\ms,\tau) = \mu_{g} \Phi_{g}(\tau) p(g) p(\ms|g)
  ~.
\end{align*}
These give:
\begin{equation}
\frac{d\rho(g,\ms,\tau)}{d\tau} = -\mu_{g} \phi_{g}(\tau)p(\ms) p(\ms|g).
\label{eq:1}
\end{equation}

Plugging Eq.~(\ref{eq:1}) into Eq.~(\ref{eq:diffyQ1}) yields:
\begin{align*}
\frac{d p(\rep|\ms,\tau)}{d\tau} \rho(g,\ms,\tau)
  & - \mu_{g} \phi_{g}(\tau)p(g)p(\ms|g) p(\rep|g,\ms,\tau) \\
  & = \sum_{\rep'\neq \rep} k_{\rep'\rightarrow \rep}(\ms) \rho(g,\ms,\tau) p(\rep'|g,\ms,\tau)
  - \frac{\phi_{g}(\tau)}{\Phi_{g}(\tau)} \rho(g,\ms,\tau) p(\rep|g,\ms,\tau)
  - k_{\rep\rightarrow\rep}(\ms) \rho(g,\ms,\tau) p(\rep|g,\ms,\tau)
  ~,
\end{align*}
where we note that:
\begin{equation*}
\mu_{g} \phi_{g}(\tau)p(g)p(\ms|g) p(\rep|g,\ms,\tau) = \frac{\phi_{g}(\tau)}{\Phi_{g}(\tau)} \rho(g,\ms,\tau) p(\rep|g,\ms,\tau)
  ~.
\end{equation*}
Hence, we are left with:
\begin{align*}
\frac{d p(\rep|g,\ms,\tau)}{d\tau}
  & = \sum_{\rep'\neq\rep} k_{\rep'\rightarrow\rep}(\ms)
  p(\rep'|g,\ms,\tau) - k_{\rep\rightarrow\rep}(\ms) p(\rep|g,\ms,\tau)
  ~.
\end{align*}

We can summarize this ordinary differential equation in matrix-vector notation
as follows. Let $\vec{v}(g,\ms,\tau)$ be the vector:
\begin{align*}
\vec{v}(g,\ms,\tau)
  := \begin{pmatrix}
  p(\rep_1|g,\ms,\tau) \\ \vdots \\ p(\rep_{|\ChanAlph|}|g,\ms,\tau)
  \end{pmatrix}
  ~.
\end{align*}
We have:
\begin{align*}
\frac{d\vec{v}}{d\tau} = M(\ms)\vec{v}
  ~.
\end{align*}
The solution to the equation above is:
\begin{align}
\vec{v}(g,\ms,\tau) = e^{M(\ms)\tau} \vec{v}(g,\ms,0)
  ~.
\label{eq:5}
\end{align}
The structure of $M(\ms)$ guarantees that probability is conserved, as long as $1^{\top}\vec{v}(g,\ms,0)=1$ for all $\ms\in\MeasAlphabet$.

Our next task is to find expressions for $\vec{v}(g,\ms,0)$.  We do this by
considering Eq.~(\ref{eq:main}) in the limit that $\tau<\Delta t$. More
straightforwardly, we consider the equation:
\begin{align}
\rho((g,\ms,0),\rep)
  & = \sum_{g',\ms'} \int_0^{\infty} d\tau
  ~\frac{\phi_{g'}(\tau)}{\Phi_{g'}(\tau)}
  \delta_{g,\epsilon^+(g',\ms')}
  p(\ms|g) \rho((g',\ms',\tau),\rep)
  ~,
\label{eq:2}
\end{align}
which is based on the following logic. For probability to flow into
$\rho((g,\ms,0),\rep)$ from $\rho((g',\ms',\tau),\rep')$, we need the dwell
time for symbol $\ms'$ to be exactly $\tau$ and for $\rep'=\rep$. (The latter
comes from the unlikelihood of switching both channel state and input symbol at
the same time.) Again decomposing:
\begin{align}
\rho((g',\ms',\tau),\rep)
  & = p(\rep|g',\ms',\tau) \rho(g',\ms',\tau) \nonumber \\
  & = \mu_{g'} \Phi_{g'}(\tau) p(g')p(\ms'|g') p(\rep|g',\ms',\tau)
\label{eq:3}
\end{align}
and, thus, as a special case:
\begin{align}
\rho((g,\ms,0),\rep) = p(\rep|g,\ms,0) p(g) p(\ms|g) \mu_{g}
  ~.
\label{eq:4}
\end{align}
Plugging both Eqs.~(\ref{eq:3}) and (\ref{eq:4}) into Eq.~(\ref{eq:2}), we
find:
\begin{align*}
\mu_{g} p(g) p(\ms|g) p(\rep|g,\ms,0)
  & = \sum_{g',\ms'} \int_0^{\infty}
  \mu_{g'} p(g') p(\ms'|g') \phi_{g'}(\tau)
  \delta_{g,\epsilon^+(g',\ms')} p(\ms|g) p(\rep|g',\ms',\tau) d\tau \\
\mu_{g} p(g) p(\rep|g,\ms,0)
  & = \sum_{g',\ms'} \int_0^{\infty}
  \mu_{g'} p(g') p(\ms'|g') \phi_{g'}(\tau)
  \delta_{g,\epsilon^+(g',\ms')} p(\rep|g',\ms',\tau) d\tau
  ~.
\end{align*}
Using Eq.~(\ref{eq:5}), we see that $p(\rep|g',\ms',\tau) =
\left(e^{M(\ms')\tau} \vec{v}(g',\ms',0)\right)_{\rep}$ and $p(\rep|g,\ms,0) =
\left(\vec{v}(g,\ms,0)\right)_{\rep}$. So, we have:
\begin{align*}
\mu_{g} p(g) \vec{v}(g,\ms,0)
  = \sum_{g',\ms'} \mu_{g'} \delta_{g,\epsilon^+(g',\ms')}
  p(g') p(\ms'|g')
  \left(\int_0^{\infty} \phi_{g'}(\tau) e^{M(\ms')\tau} d\tau \right)
  \vec{v}(g',\ms',0)
  ~.
\end{align*}
If we form the composite vector $\vec{V}$ as:
\begin{align*}
\vec{U}
  & = \begin{pmatrix}
  \vec{u}(g_1,\ms_1) \\
  \vec{u}(g_1,\ms_2) \\
  \vdots \\
  \vec{u}(g_{|\mathcal{G}|},\ms_{|\MeasAlphabet|})
  \end{pmatrix} \\
  & := \begin{pmatrix}
  \mu_{g_1} p(g_1) \vec{v}(g_1,\ms_1,0) \\
  \vdots \\
  \mu_{g_{|\mathcal{G}|}} p(g_{|\mathcal{G}|})
  \vec{v}(g_{|\mathcal{G}|},\ms_{|\MeasAlphabet|},0)
  \end{pmatrix}
\end{align*}
and the matrix (written in block form) as:
\begin{align*}
\bf{C}
  := \begin{pmatrix}
  C_{(g_1,\ms_1)\rightarrow (g_1,\ms_1)} & C_{(g_1,\ms_2)\rightarrow (g_1,\ms_1)} & \ldots \\
  C_{(g_1,\ms_1)\rightarrow (g_1,\ms_2)} & C_{(g_1,\ms_2)\rightarrow (g_1,\ms_2)} & \ldots \\
  \vdots & \vdots & \ddots
  \end{pmatrix}
  ~,
\end{align*}
with:
\begin{align*}
C_{(g',\ms')\rightarrow (g,\ms)}
  = \delta_{g,\epsilon^+(g',\ms')} p(\ms'|g')
  \int_0^{\infty} \phi_{g'}(t) e^{M(\ms') t} dt
  ~,
\end{align*}
we then have:
\begin{align}
\vec{U} = \text{eig}_1(\bf{C})
  ~.
\label{eq:init_conds}
\end{align}
Finally, we must normalize $\vec{u}(\ms)$ appropriately. We do this by
recalling that $1^{\top}\vec{v}(g,\ms,0)=1$, since $\vec{v}(g,\ms,0)$ is a
vector of probabilities. Then we have:
\begin{align*}
\vec{u}(g,\ms)
  \rightarrow \frac{\vec{u}(g,\ms)}{1^{\top}\vec{u}(g,\ms)} \mu_{g} p(g)
  ~.
\end{align*}
for each $g,\ms$.

To calculate predictive metrics---i.e., $I_{mem}$ and $I_{fut}$---we need
$p(\ms,\rep)$ and $p(\rep,\pst)$. The former is a marginalization of
$p(\fst,\rep)$ that we just calculated. The second can be calculated via:
\begin{align*}
p(\pst,\rep) = \sum_{\fst} p(\pst|\fst) p(\rep,\fst)
  ~,
\end{align*}
where
\begin{align*}
p(\pst|\fst) & = p((g_-,\ms_-,\tau_-)|(g_+,\ms_+,\tau_+)) \\
  & = \delta_{\ms_+,\ms_-} p(g_-|g_+,\ms_+)
  \mu_{g_+} \phi_{g_+}(\tau_+ +\tau_-)
  ~.
\end{align*}
Hence, we turn our attention to dissipative metrics.

For calculation of dissipative metrics, we only need:
\begin{align*}
\frac{\delta p}{\delta t}
  = \lim_{\Delta t\rightarrow 0}
  \frac{\Prob(\MS_{t+\Delta t}=\ms,\Rep_t=\rep) - \Prob(\MS_t=\ms,\Rep_t=\rep)}{\Delta t}
  ~.
\end{align*}
Moreover, we can use the Markov chain
$\Rep_t\rightarrow\FSt_t\rightarrow\MS_{t+\Delta t}$ to compute it:
\begin{align*}
\Prob(\MS_{t+\Delta t}=\ms,\Rep_t=\rep)
  = \sum_{\fst}
  \Prob(\MS_{t+\Delta t}=\ms|\FSt_t=\fst) \Prob(\Rep_t=\rep,\FSt_t=\fst)
  ~.
\end{align*}
We have:
\begin{align*}
\Prob(\MS_{t+\Delta t}=\ms|\FSt_t=\fst)
  & = \Prob(\MS_{t+\Delta t}=\ms|\FSt_t=(g',\ms',\tau')) \\
  & = \begin{cases}
  \frac{\Phi_{g'}(\tau'+\Delta t)}{\Phi_{g'}(\tau')}
    & \ms=\ms' \\
  \frac{\phi_{g'}(\tau')}{\Phi_{g'}(\tau')} p(\ms|\epsilon^+(g',\ms')) \Delta t
    & \ms\neq\ms'
  \end{cases}
  ~.
\end{align*}
This, combined with $p(\fst,\rep)$, gives:
\begin{align*}
\Prob(\MS_{t+\Delta t}=\ms,\Rep_t=\rep)
  & = \sum_{g',\ms'\neq\ms} \int d\tau'~\rho((g',\ms',\tau'),\rep) \frac{\phi_{g'}(\tau')}{\Phi_{g'}(\tau')} \Delta t~p(\ms|\epsilon^+(g',\ms')) \\
  & \qquad + \sum_{g'} \int d\tau' \frac{\Phi_{g'}(\tau'+\Delta t)}{\Phi_{g'}(\tau')} \rho((g',\ms',\tau'),\rep) \\
  & = \Prob(\MS_{t}=\ms,\Rep_t=\rep) + \Delta t \Big( \sum_{g',\ms'\neq\ms} \int d\tau' p(\ms|\epsilon^+(g',\ms')) \frac{\phi_{g'}(\tau')}{\Phi_{g'}(\tau')} \rho((g',\ms',\tau'),\rep) \nonumber \\
  & \qquad -\sum_{g'} \int d\tau' \frac{\phi_{g'}(\tau')}{\Phi_{g'}(\tau')} \rho((g',\ms,\tau'),\rep)\Big)
  ~,
\end{align*}
correct to $O(\Delta t)$. Recalling that:
\begin{align*}
\rho((g',\ms',\tau'),\rep)
  & = \rho(g',\ms',\tau')p(\rep|g',\ms',\tau') \\
  & = p(\ms'|g') \Phi_{g'}(\tau')
  \left(e^{M(\ms')\tau'}\vec{u}(g',\ms')\right)_{\rep}
  ~,
\end{align*}
gives:
\begin{align}
\frac{\delta p}{\delta t}
  & = \lim_{\Delta t\rightarrow 0}
  \frac{\Prob(\MS_{t+\Delta t}=\ms,\Rep_t=\rep)
  -\Prob(\MS_{t}=\ms,\Rep_t=\rep)}{\Delta t} \nonumber \\
  & = \sum_{g',\ms'\neq\ms} \int d\tau' p(\ms|\epsilon^+(g',\ms')) 
  \frac{\phi_{g'}(\tau')}{\Phi_{g'}(\tau')} \rho((g',\ms',\tau'),\rep) 
  -\sum_{g'} \int d\tau' \frac{\phi_{g'}(\tau')}{\Phi_{g'}(\tau')}
  \rho((g',\ms,\tau'),\rep) \\
  & = \sum_{g',\ms'\neq\ms} \int d\tau' ~p(\ms|\epsilon^+(g',\ms')) p(\ms'|g') \phi_{g'}(\tau') \left(e^{M(\ms')\tau'}\vec{u}(g',\ms')\right)_{\rep} \nonumber \\
  & - \sum_{g'} \int d\tau'~p(\ms|g') \phi_{g'}(\tau')
  \left(e^{M(\ms)\tau'}\vec{u}(g',\ms)\right)_{\rep}
  ~.
\label{eq:deltap_deltat1}
\end{align}
From this, Eqs.~(\ref{eq:Inps}) and (\ref{eq:betaP}) can be used to calculate
$\Inpdot$ and $\beta P$.

\section{Specialization to semi-Markov input}

Up to this point, we wrote expressions for the general case of unifilar hidden
semi-Markov input. We now specialize to the semi-Markov input case. A great
simplification ensues: hidden states $g$ are the current emitted symbols $\ms$.
Recall that, in an abuse of notation, $q(\ms|\ms')$ is now the probability of
observing symbol $\ms$ after seeing symbol $\ms'$.

Hence, forward-time causal states are given by the pair $(\ms,\tau)$. The
analog of Eq.~(\ref{eq:5}) is:
\begin{align*}
\vec{p}(\rep|\ms,\tau) = e^{M(\ms)\tau} \vec{p}(\rep|\ms,0)
  ~,
\end{align*}
and we define vectors:
\begin{align*}
\vec{u}(\ms) := \mu_{\ms} p(\ms) \vec{p}(\rep|\ms,0)
  ~.
\end{align*}
The large vector:
\begin{align*}
\vec{U} := \begin{pmatrix}
\vec{u}(\ms_1) \\ \vdots \\ \vec{u}(\ms_{|\MeasAlphabet})
  \end{pmatrix}
\end{align*}
is the eigenvector $\text{eig}_1(\textbf{C})$ of eigenvalue $1$ of the matrix:
\begin{align*}
\textbf{C} = \begin{pmatrix}
  0 & q(\ms_1|\ms_2) \int_0^{\infty} \phi_{\ms_2}(\tau) e^{M(\ms_2)\tau}d\tau \\
  \ldots \\
  q(\ms_2|\ms_1) \int_0^{\infty} \phi_{\ms_1}(\tau) e^{M(\ms_1)\tau}d\tau \\
  \ldots \\
  \vdots & \vdots & \ddots
  \end{pmatrix}
  `,
\end{align*}
where normalization requires $1^{\top} \vec{u}(\ms) = \mu_{\ms} p(\ms)$.

We continue by finding $p(\rep)$, since from this we obtain $\H[\Rep]$. We do
this via straightforward marginalization:
\begin{align*}
p(\rep) & = \sum_{\fst} \rho(\fst,\rep) = \sum_{\fst} p(\rep|\fst) \rho(\fst) \\
  & = \sum_{\ms} \int_0^{\infty}~p(\rep|\ms,\tau) \rho(\ms,\tau)~d\tau \\
  & = \sum_{\ms} \int_0^{\infty}
  \left(e^{M(\ms)\tau} \vec{v}(\ms,0)\right)_{\rep}
  \mu_{\ms} p(\ms) \Phi_{\ms}(\tau) d\tau \\
  & = \sum_{\ms} \left(\left(\int_0^{\infty}
  e^{M(\ms)\tau} \Phi_{\ms}(\tau) d\tau\right)\vec{u}(\ms)\right)_{\rep}
  ~.
\end{align*}
This implies that:
\begin{align*}
\vec{p}(\rep)
  = \sum_{\ms}
  \left(\int_0^{\infty} e^{M(\ms)\tau} \Phi_{\ms}(\tau) d\tau\right)
  \vec{u}(\ms)
   ~.
\end{align*}
From earlier, recall that $\vec{u}(\ms) := \mu_{\ms} p(\ms) \vec{p}(\rep|\ms,0)$.

Next, we aim to find $p(\ms,\rep)$, again via marginalization:
\begin{align}
p(\ms,\rep)
  & = \int_0^{\infty} \rho((\ms,\tau),\rep) d\tau \nonumber \\
  & = \int_0^{\infty} \mu_{\ms} p(\ms) \Phi_{\ms}(\tau) p(\rep|\ms,\tau) d\tau
  \nonumber \\
  & = \int_0^{\infty} \mu_{\ms} p(\ms) \Phi_{\ms}(\tau)
  \left(e^{M(\ms)\tau} \vec{v}(\ms,0)\right)_{\rep} d\tau \nonumber \\
  & = \left(\left(\int_0^{\infty} e^{M(\ms)\tau}
  \Phi_{\ms}(\tau)d\tau\right) \vec{u}(\ms) \right)_{\rep}
  ~.
\label{eq:pmsrep}
\end{align}
From the joint distribution $p(\ms,\rep)$, we easily numerically obtain
$\I[\MS;\Rep]$, since $|\MeasAlphabet| < \infty$ and $|\ChanAlph| <\infty$.

For notational ease, we introduced $\mathcal{T}_t$ in this section as the
random variable for the time since last symbol, whose realization is $\tau$.
Finally, we require $p(\rep|\pst)$ to calculate $\H[\Rep|\PSt]$, which we can
then combine with $\H[\Rep]$ to get an estimate for $\Ifut$. We utilize the
Markov chain $\Rep\rightarrow\FSt\rightarrow\PSt$, as stated earlier, and so
have:
\begin{align*}
p(\rep|\pst) & = \sum_{\fst} \rho(\rep,\fst|\pst) \\
  & = \sum_{\fst} p(\rep|\fst,\pst) \rho(\fst|\pst) \\
  & = \sum_{\fst} p(\rep|\fst) \rho(\fst|\pst)
  ~.
\end{align*}
Eq.~(\ref{eq:5}) gives us $p(\rep|\fst)$ as:
\begin{align*}
p(\rep|\fst) & = p(\rep|\ms_+,\tau_+) \\
   & = \left(e^{M(\ms_+)\tau_+} \vec{v}(\ms_+,0)\right)_{\rep}
\end{align*}
and Eq.~(\ref{eq:cond_pdf}) gives us $\rho(\fst|\pst)$ after some manipulation:
\begin{align*}
\rho(\fst|\pst)
  & = \rho((\ms_+,\tau_+)|(\ms_-,\tau_-)) \\
  & = \delta_{\ms_+,\ms_-}
  \frac{\phi_{\ms_-}(\tau_+ + \tau_-)}{\Phi_{\ms_-}(\tau_-)}
  ~.
\end{align*}
Combining the two equations gives:
\begin{align*}
p(\rep|\ms_-,\tau_-)
  & = \sum_{\ms_+} \int_0^{\infty}~\delta_{\ms_+,\ms_-}
  \frac{\phi_{\ms_-}(\tau_+ + \tau_-)}{\Phi_{\ms_-}(\tau_-)}
  \left(e^{M(\ms_+)\tau_+} \vec{v}(\ms_+,0)\right)_{\rep}~d\tau_+ \\
  & = \frac{1}{\Phi_{\ms_-}(\tau_-)}
  \left(\left(\int_0^{\infty} \phi_{\ms_-}(\tau_+ + \tau_-)
  e^{M(\ms_-)\tau_+} d\tau_+ \right)
  \vec{v}(\ms_-,0)\right)_{\rep}
  ~.
\end{align*}
From this conditional distribution, we compute $\H[\Rep|\PSt=\pst]$, and so
$\H[\Rep|\PSt]=\langle \H[\Rep|\PSt=\pst]\rangle_{\rho(\pst)}$. In more detail,
define:
\begin{align*}
D_{\ms}(\tau) := \int_0^{\infty} \phi_{\ms}(\tau+s) e^{M(\ms)s} ds
  ~,
\end{align*}
and we have:
\begin{align*}
\vec{p}(\rep|\ms_-,\tau_-)
  = D_{\ms_-}(\tau_-)\vec{u}(\ms_-)/\mu_{\ms_-}p(\ms_-)\Phi_{\ms_-}(\tau_-)
  ~.
\end{align*}
This conditional distribution gives:
\begin{align*}
H[\Rep|\MS_-=\ms_-,\Tau_-=\tau_-]
  & = -\sum_{\rep} p(\rep|\ms_-,\tau_-)\log p(\rep|\ms_-,\tau_-) \\
  & = - 1^{\top}
  \left(\frac{D_{\ms_-}(\tau_-)\vec{u}(\ms_-)}{\mu_{\ms_-}p(\ms_-)\Phi_{\ms_-}(\tau_-)}
  \log \left(
  \frac{D_{\ms_-}(\tau_-)\vec{u}(\ms_-)}{\mu_{\ms_-}p(\ms_-)\Phi_{\ms_-}(\tau_-)}
  \right)\right) \\
  & = -\frac{1}{\mu_{\ms_-}p(\ms_-)\Phi_{\ms_-}(\tau_-)}
  \Big(1^{\top}
  \left((D_{\ms_-}(\tau_-)\vec{u}(\ms_-))
  \log (D_{\ms_-}(\tau_-)\vec{u}(\ms_-))\right) \\
  & \qquad -1^{\top} (D_{\ms_-}(\tau_-)\vec{u}(\ms_-))
  \log (\mu_{\ms_-}p(\ms_-)\Phi_{\ms_-}(\tau_-))\Big)
  ~.
\end{align*}
We recognize the factor $\mu_{\ms_-}p(\ms_-)\Phi_{\ms_-}(\tau_-)$ as
$\rho(\ms_-,\tau_-)$ and so we find that:
\begin{align*}
H[\Rep|\MS_-,\Tau_-]
  & = \sum_{\ms_-} \int_0^{\infty}~\rho(\ms_-,\tau_-)
  \ H[\Rep|\MS_-=\ms_-,\Tau_-=\tau_-] d\tau_- \\
  & = -\int_0^{\infty}
  \left(\sum_{\ms_-} 1^{\top} \left((D_{\ms_-}(\tau_-)\vec{u}(\ms_-))\log (D_{\ms_-}(\tau_-)\vec{u}(\ms_-))
  \right)\right) d\tau_- \\
  & \qquad + \int_0^{\infty}
  \left(\sum_{\ms_-} 1^{\top} D_{\ms_-}(\tau_-)
  \vec{u}(\ms_-)\log (\mu_{\ms_-}p(\ms_-)\Phi_{\ms_-}(\tau_-))\right) d\tau_-
  ~.
\end{align*}
This, combined with earlier formula for $\H[\Rep]$, gives $\Ifut$.


Finally, we wish to find an expression for the nonpredictive information rate
$\Inpdot$. We review the somewhat compact derivation of $\frac{\delta
p}{\delta t}$ in the more general case, specialized for semi-Markov input.
This requires finding an expression for $\Prob(\Rep_t=\rep,\MS_{t+\Delta
t}=\ms)$ as an expansion in $\Delta t$. We start as usual:
\begin{align*}
\Prob(\Rep_t=\rep,\MS_{t+\Delta t}=\ms) = \sum_{\ms'} \int_0^{\infty} \Prob(\Rep_t=\rep,\MS_{t+\Delta t}=\ms,\MS_t=\ms',\Tau_t=\tau) d\tau 
\end{align*}
and utilize the Markov chain $\Rep_t\rightarrow\FSt_t\rightarrow\MS_{t+\Delta
t}$, giving:
\begin{align}
\Prob(\Rep_t=\rep,\MS_{t+\Delta t}=\ms)
  = \sum_{\ms'} \int_0^{\infty}
  \Prob(\Rep_t=\rep|\MS_t=\ms',\Tau_t=\tau) \Prob(\MS_{t+\Delta t}=\ms|\MS_t=\ms',\Tau_t=\tau) \rho(\ms',\tau)
  d\tau
  ~.
\label{eq:10}
\end{align}
We have $\Prob(\Rep_t=\rep|\MS_t=\ms,\mathcal{T}_t=\tau)$ from
Eq.~(\ref{eq:5}). So, we turn our attention to finding $\Prob(\MS_{t+\Delta
t}=\ms|\MS_t=\ms',\mathcal{T}_t=\tau)$. Some thought reveals that:
\begin{align}
\Prob(\MS_{t+\Delta t}=\ms|\MS_t=\ms',\Tau_t=\tau)
  = \begin{cases}
  q(\ms|\ms')\frac{\phi_{\ms'}(\tau)}{\Phi_{\ms'}(\tau)} \Delta t & \ms\neq\ms' \\
  \frac{\Phi_{\ms'}(\tau+\Delta t)}{\Phi_{\ms'}(\tau)} & \ms=\ms'
  \end{cases}
  ~,
\label{eq:11}
\end{align}
plus corrections of $O(\Delta t^2)$. We substitute Eq.~(\ref{eq:11}) into
Eq.~(\ref{eq:10}) to get:
\begin{align*}
\Prob(\Rep_t=\rep,\MS_{t+\Delta t}=\ms)
  & = \left( \sum_{\ms'\neq \ms} \int_0^{\infty}
  \Prob(\Rep_t=\rep|\MS_t=\ms',\Tau_t=\tau) q(\ms|\ms')
  \frac{\phi_{\ms'}(\tau)}{\Phi_{\ms'}(\tau)}
  \rho(\ms',\tau)d\tau\right) \Delta t \\
  & \qquad + \int_0^{\infty}
  \Prob(\Rep_t=\rep|\MS_t=\ms,\Tau_t=\tau)
  \frac{\Phi_{\ms}(\tau+\Delta t)}{\Phi_{\ms}(\tau)}
  \rho(\ms,\tau)d\tau
  ~,
\end{align*}
plus corrections of $O(\Delta t^2)$. Recalling:
\begin{align*}
\frac{\Phi_{\ms}(\tau+\Delta t)}{\Phi_{\ms}(\tau)}
  = 1-\frac{\phi_{\ms}(\tau)}{\Phi_{\ms}(\tau)} \Delta t
  ~,
\end{align*}
plus corrections of $O(\Delta t^2)$, we simplify further:
\begin{align*}
\Prob(\Rep_t=\rep,\MS_{t+\Delta t}=\ms)
  & = \left( \sum_{\ms'\neq \ms} \int_0^{\infty}
  \Prob(\Rep_t=\rep|\MS_t=\ms',\Tau_t=\tau) q(\ms|\ms')
  \frac{\phi_{\ms'}(\tau)}{\Phi_{\ms'}(\tau)}
  \rho(\ms',\tau)d\tau\right) \Delta t \\
  & \qquad + \int_0^{\infty}
  \Prob(\Rep_t=\rep|\MS_t=\ms,\Tau_t=\tau) \rho(\ms,\tau)d\tau \\
  & \qquad -\left( \int_0^{\infty} \Prob(\Rep_t=\rep|\MS_t=\ms,\Tau_t=\tau)
  \frac{\phi_{\ms}(\tau)}{\Phi_{\ms}(\tau)}
  \rho(\ms,\tau)d\tau \right)\Delta t
  ~,
\end{align*}
plus $O(\Delta t^2)$ corrections. We notice that:
\begin{align*}
\int_0^{\infty} \Prob(\Rep_t=\rep|\MS_t=\ms,\Tau_t=\tau) \rho(\ms,\tau)d\tau
  = \Prob(\Rep_t=\rep,\MS_t=\ms)
  ~,
\end{align*}
so that:
\begin{align*}
\lim_{\Delta t\rightarrow 0}
  \frac{\Prob(\Rep_t=\rep,\MS_{t+\Delta t}=\ms)-\Prob(\Rep_t=\rep,\MS_t=\ms)}{\Delta t}
  & = \sum_{\ms'\neq \ms} \int_0^{\infty}
  \Prob(\Rep_t=\rep|\MS_t=\ms',\Tau_t=\tau) q(\ms|\ms')
  \frac{\phi_{\ms'}(\tau)}{\Phi_{\ms'}(\tau)}  \rho(\ms',\tau)d\tau \\
  & \qquad -\int_0^{\infty} \Prob(\Rep_t=\rep|\MS_t=\ms,\Tau_t=\tau) 
  \frac{\phi_{\ms}(\tau)}{\Phi_{\ms}(\tau)}  \rho(\ms,\tau)d\tau
  ~.
\end{align*}
Substituting Eqs.~(\ref{eq:5}) and (\ref{eq:marg_pdf}) into the above
expressions yields:
\begin{align*}
\sum_{\ms'\neq \ms} \int_0^{\infty}
  \Prob(\Rep_t=\rep|\MS_t=\ms',\Tau_t=\tau) q(\ms|\ms')
  \frac{\phi_{\ms'}(\tau)}{\Phi_{\ms'}(\tau)}
  \rho(\ms',\tau)d\tau
  & = \sum_{\ms'} q(\ms|\ms') 
  \left(
  \left( \int_0^{\infty} \phi_{\ms'}(\tau) e^{M(\ms')\tau} d\tau \right)
  \vec{u}(\ms')\right)_{\rep}
\end{align*}
and:
\begin{align*}
\int_0^{\infty} \Prob(\Rep_t=\rep|\MS_t=\ms,\Tau_t=\tau)
  \frac{\phi_{\ms}(\tau)}{\Phi_{\ms}(\tau)} \rho(\ms,\tau)d\tau
  & = \left
  (\left(\int_0^{\infty} \phi_{\ms}(\tau) e^{M(\ms)\tau} d\tau \right)
  \vec{u}(\ms)\right)_{\rep}
  ~,
\end{align*}
so that we have:
\begin{align*}
\lim_{\Delta t\rightarrow 0} &
  \frac{\Prob(\Rep_t=\rep,\MS_{t+\Delta
  t}=\ms) -\Prob(\Rep_t=\rep,\MS_t=\ms)}{\Delta t} \\
  & \qquad\qquad\qquad = \Big( \sum_{\ms'} q(\ms|\ms')
  \left(\int_0^{\infty} \phi_{\ms'}(\tau) e^{M(\ms')\tau} d\tau \right)
  \vec{u}(\ms') 
  - \left(\int_0^{\infty} \phi_{\ms}(\tau) e^{M(\ms)\tau} d\tau \right)
  \vec{u}(\ms)\Big)_{\rep}
  ~.
\end{align*}
For notational ease, denote the lefthand side as $\delta p(\ms,\rep)
/ \delta t$. The nonpredictive information rate is given by:
\begin{align*}
\Inpdot & = \lim_{\Delta t\rightarrow 0}
  \frac{\I[\MS_t;\Rep_t]-\I[\MS_{t+\Delta t};\Rep_t]}{\Delta t} \\
  & = \lim_{\Delta t\rightarrow 0}
  \frac{\left(\H[\MS_t]+H[\Rep_t] - \H[\MS_t,\Rep_t]\right)
  - \left(\H[\MS_{t+\Delta t}] + \H[\Rep_t]
  - \H[\MS_{t+\Delta t},\Rep_t]\right)}{\Delta t} \\
  & = \lim_{\Delta t\rightarrow 0}
  \frac{\H[\MS_{t+\Delta t},\Rep_t] - \H[\MS_t,\Rep_t]}{\Delta t}
  ~,
\end{align*}
where we utilize stationarity to assert $\H[\MS_t]=\H[\MS_{t+\Delta t}]$.
Then, correct to $O(\Delta t)$, we have:
\begin{align*}
H[\MS_{t+\Delta t},\Rep_t]
  & = - \sum_{\ms,\rep}
  \left(p(\ms,\rep) + \frac{\delta p(\ms,\rep)}{\delta t}\Delta t\right)
  \log \left(p(\ms,\rep) + \frac{\delta p(\ms,\rep)}{\delta t}\Delta t\right) \\
  & = -\sum_{\ms,\rep} p(\ms,\rep) \log p(\ms,\rep)
  - \sum_{\ms,\rep}p(\ms,\rep)
  \frac{\delta p(\ms,\rep)/\delta t}{p(\ms,\rep)}~\Delta t
  - \sum_{\ms,\rep}
  \frac{\delta p(\ms,\rep)}{\delta t} \log p(\ms,\rep) \Delta t \\
  & = \H[\MS_t;\Rep_t]
  - \sum_{\ms,\rep} \frac{\delta p(\ms,\rep)}{\delta t} \log p(\ms,\rep)
  \Delta t
  ~,
\end{align*}
which implies:
\begin{align*}
\Inpdot & = \sum_{\ms,\rep}
  \frac{\delta p(\ms,\rep)}{\delta t} \log p(\ms,\rep)
  ~,
\end{align*}
with:
\begin{align}
\frac{\delta p(\ms,\rep)}{\delta t}
  & = \left( \sum_{\ms'} q(\ms|\ms')
  \left(\int_0^{\infty} \Phi_{\ms'}(\tau) e^{M(\ms')\tau} d\tau \right)
  \vec{u}(\ms')
  - \left(\int_0^{\infty} \phi_{\ms}(\tau) e^{M(\ms)\tau} d\tau \right)
  \vec{u}(\ms)\right)_{\rep}
\label{eq:deltap_deltat2}
\end{align}
and $p(\ms,\rep)$ given in Eq.~(\ref{eq:pmsrep}).

\end{widetext}

\end{document}